\documentclass[aps,twocolumn,amssymb,prb]{revtex4}

\usepackage{amsfonts}
\usepackage{subfigure}
\usepackage{epsfig}
\usepackage{float}

\begin{document}

\renewcommand{\thefootnote}{\fnsymbol{footnote}}
\renewcommand{\theequation}{\arabic{section}.\arabic{equation}}

\title{Inverse Optimization Techniques for Targeted Self-Assembly}

\author{Salvatore Torquato}
\email[]{E-mail: torquato@princeton.edu}

\affiliation{Department of Chemistry, Princeton Center for Theoretical Science, Princeton Institute for the Science and Technology of Materials, and
Program in Applied and Computational Mathematics, Princeton University, Princeton, New Jersey 08544, USA; \\and School of Natural Sciences, 
Institute for Advanced Study, Princeton, New Jersey, 08540, USA }

\date{\today}

\begin{abstract}

This article reviews recent inverse statistical-mechanical
methodologies that we have devised to optimize interaction potentials in soft matter
systems  that correspond to  stable ``target"
structures. We are interested in finding the interaction potential,
not necessarily pairwise additive or spherically symmetric,
 that stabilizes a targeted many-body system by generally incorporating
complete configurational information.
Unlike previous work, our primary interest is in the possible
many-body structures that may be generated, some of which may include
interesting but known structures, while others may represent entirely new
structural motifs. Soft matter systems, such as colloids
and polymers,  offer a versatile means of realizing the optimized interactions.
It is shown that these inverse approaches hold
great promise for controlling self-assembly to a 
degree  that surpasses the less-than-optimal path that nature has provided.
Indeed, we envision being able to  ``tailor''
potentials that produce varying degrees of
disorder, thus extending the traditional idea of self-assembly to
incorporate  both amorphous and crystalline structures as well
as quasicrystals. The notion
of tailoring potentials that correspond to targeted structures is
motivated by the rich fundamental statistical-mechanical
issues and questions offered by
this fascinating inverse  problem as well
as our recent ability to identify structures that
have optimal bulk properties or desirable performance
characteristics. Recent results have already led to  
a deeper basic understanding of the mathematical relationship between
the collective structural behavior of many-body systems and their interactions,
as well as optimized potentials that enable self-assembly of
ordered and disordered particle configurations with novel structural and
bulk properties. 
\end{abstract}

\maketitle


\section{Introduction}
\label{intro}

The term ``self-assembly"  typically describes processes in
which entities (atoms, molecules, aggregates of molecules, etc.)
spontaneously arrange themselves into a larger ordered and functioning
structure. Biology offers wonderful examples, including the spontaneous
formation of the DNA double helix from two complementary oligonucleotide chains, the
formation of lipid bilayers to produce membranes, and the folding
of proteins into a biologically active state. 

On the synthetic side, molecular self-assembly is a potentially powerful 
method to fabricate nanostructures as an alternative to nanolithography.
For example, it has been demonstrated that
intricate two-dimensional structures can emerge by the placement of organic
molecules onto inorganic surfaces. \cite{Whi90}
Block copolymers can self-assemble into ordered
arrays that have possible use as photonic band-gap materials. \cite{Jen99}
Self-assembly based on contact electrification
seems to be a powerful means to organize macroscopic dielectric
particles of various shapes into extended, ordered structures. \cite{Gr03}
Highly robust self-assembly of unique, small clusters
of microspheres that can themselves be used for self-assembly of more
complex architectures has been demonstrated. \cite{Mano03}
It has been shown that gold nanowires can be assembled by functionalizing nanoparticles
with organic molecules. \cite{Stell04}  DNA-mediated
assembly of micrometer-size polystyrene particles in solution
could enable one to build complex structures starting from a
mesoscale template or seed followed by self-assembly. \cite{Chaik05} These examples offer glimpses into the materials science
of the future -- devising building blocks with specific interactions
that can self-organize on a larger set of length scales.

This is an emerging field with a wealth of experimental data that 
has been supported theoretically and computationally using 
the ``forward'' approach of statistical mechanics. 
\cite{Ha86,Al87,Fr96,Wa99,La00,Zh04,Ar04,Hyn06,Gl07,Bia08} Such an
approach has generated a long and insightful tradition. The forward approach identifies a 
known material system that possesses scientific and/or technological interest, creates
a manageable approximation to the interparticle interactions that operate
in that material, and exploits simulation and analytical methods
to predict non-obvious details concerning structural, thermodynamic and kinetic
features of the system.

In the last several years, {\it inverse} statistical-mechanical methods 
have been devised that find optimized interactions that most robustly 
and spontaneously lead to a targeted many-particle configuration of the system
for a wide range of conditions. 
\cite{Uc04b,Re05,Re06a,Re06b,Uc06b,Re07a,Re07b,Re08a,Re08b,Re08c,Ba08}
This article reviews these nascent developments as well as other closely
related inverse realizability problems that we have introduced, \cite{St01a,Sa02,To02c,Cr03,St04,St05,Uc06a,To06b,To06d,Sc08} all of which are 
solved using various optimization techniques.

Results produced by these inverse approaches have already led to
a deeper fundamental understanding of the
mathematical relationship between
the collective structural behavior
of many-body systems and the interactions: a basic
problem in materials science and condensed matter theory.
As will be shown, such methodologies  hold
great promise to control self-assembly in many-particle systems to a degree 
that surpasses the less-than-optimal path that nature has provided. 
Indeed, employing such inverse optimization methods, we envision being able to  ``tailor''
potentials that produce varying degrees of
disorder, thus extending the traditional idea of self-assembly to
incorporate  both amorphous and crystalline structures as well
as quasicrystals. 

Output from these optimization techniques could then be applied 
to create {\it de novo}  colloidal particles or polymer systems with
interactions that yield these structures at the nanoscopic  and microscopic
length scales. Thus, this work has important implications
for the future synthesis of novel materials.

Colloidal particles suspended in solution provide
an ideal experimental testbed to realize the optimized
potentials, since both repulsive and
attractive interactions can be tuned ({\it e.g.}, via particle surface
modification or the addition of electrolytes). \cite{Ru89,Mu96,Chaik01,Mano03,Chaik05,Hyn06,Chaik07,Ye07}
and therefore offer a panoply of {\it possible} potentials that
far extends the range offered by molecular systems.
Effective pair interactions in colloids can contain hard-core,
charge dispersion (van der Waals), dipole-dipole 
(electric- or magnetic-field induced \cite{Ye07}), screened-Coulombic (Yukawa), and 
short-ranged attractive depletion contributions.  
Polymer systems also offer a versatile means of
realizing optimized soft interactions. \cite{La00,Ml06}

The idea of {\it tailoring} potentials  to generate targeted structures is
motivated by the rich array of fundamental issues and questions offered by 
this fascinating inverse statistical-mechanical problem as well
as our recent ability to identify the structures that
have optimal bulk properties or desirable performance
characteristics.  The latter includes novel crystal \cite{Ho90,Ka05,Si08}
and quasicrystal \cite{Man05,Re08b} structures for 
photonic band-gap applications, materials with
negative or vanishing 
thermal expansion coefficients, \cite{Si96,Mol,Gi97c} materials
with negative Poisson ratios, \cite{La87,Mi92,Si96-1,Gi97,Wei98,Xu99,Re08c}
materials with optimal or novel transport and mechanical properties, \cite{To98a,To98,Hy01,To02a,To02d,To04b,Ju05} mesoporous
solids for applications in catalysis, separations, sensors
and electronics, \cite{Fer99,Cha01} and  systems
characterized by entropically driven inverse freezing, \cite{Gre00,Fe03} to mention  a few examples.

The recent inverse techniques that are reviewed here differ from  
so-called ``reverse" Monte Carlo methods  \cite{Ke90,Ly95,Me00,Mc01} 
in several important respects. The latter techniques are concerned almost invariably with obtaining a  spherically symmetric pair potential 
from an experimentally observed structure factor (as measured from
scattering experiments) or real-space
pair correlation function usually for stable liquid phases or glassy
states of matter.
By contrast, our interest is  in finding the interaction potential,
not necessarily pairwise additive or spherically symmetric,
 that optimally stabilizes a targeted many-body system, which
may be a crystal, disordered or quasicrystal structure, by incorporating
structural information that is not limited to the pair correlation function
and generally accounts for complete configurational information.
Unlike previous work, our primary interest is in the possible 
many-body structures that may be generated, some of which may include
interesting but known structures, while others may represent entirely new
structural motifs. Moreover, the inverse methods described here
can be employed to find targeted structures for metastable states
as well as nonequilibrium configurations.

In Section \ref{background}, we define essential
terms and briefly review basic concepts that
are germane to the remainder of the article.
Section \ref{method1} describes inverse optimization methods
for self-assembly of crystal ground states.
In Section \ref{low-coord}, we discuss recent applications
of these methods that yield unusual crystal ground states with optimized,
nondirectional interactions, including
low-coordinated structures, chain-like arrays,
and lattices of clusters. Section \ref{method2} describes and applies inverse optimization methods
for self-assembly of disordered ground states.
In Section \ref{Duality}, new duality relations
for classical ground states are reviewed
and applied to some cases examined in the previous
sections. Section \ref{realize} discusses the pair-correlation-function
realizability problem and inverse
optimization procedures
to construct configurations with a given pair correlation
function. In Section \ref{bulk}, we discuss
inverse optimization methods to optimize interactions for targeted bulk properties
and specific applications.
Finally, in Section \ref{future}, we suggest
problems for future work and close with concluding remarks.

\section{Basic Definitions and Concepts}
\label{background}

We consider  a configuration of $N$ identical interacting particles with
coordinates ${\bf r}^N \equiv {\bf r}_1, {\bf r}_2, \cdots, {\bf r}_N$ in 
a region of volume $V$ in $d$-dimensional Euclidean space $\mathbb{R}^d$.
The coordinate ${\bf r}_i$ of the $i$th particle generally embodies both its
center-of-mass position and orientation as well as conformation if required. 
In the absence of an external field,
the classical $N$-body potential $\Phi_N$ can be decomposed into
2-body terms, 3-body etc., as follows:
\begin{equation}
\Phi_N({\bf r}^N) =   \sum_{i<j}^N
\varphi_2 ({\bf r}_i,{\bf r}_j)
+ \sum_{i<j<k}^N \varphi_3 ({\bf r}_i, {\bf r}_j, {\bf r}_k) +\cdots.
\end{equation}
Here $\varphi_n$ is the intrinsic $n$-body potential
in {\it excess} to the contributions
from $\varphi_2, \varphi_3, \cdots, \varphi_{n-1}$.

 A given many-body structure is specified by the local density $\rho(\bf r)$,
which can be expressed in terms of the particle coordinates as follows:
\begin{equation}
\rho({\bf r})= \sum_{i=1}^{N} \delta({\bf r} -{\bf r}_i),
\label{local}
\end{equation}
where $\delta({\bf r})$ is the $d$-dimensional Dirac delta function.
The $N$  particle coordinates 
${\bf r}^N$ are statistically characterized
by the ensemble (equilibrium or not) under consideration. The ensemble average
of $n$ products of the local densities at $n$ different positions
yields the standard $n$-particle correlation functions. \cite{To06b}
 For statistically homogeneous systems in a volume $V$, these correlation functions are defined so that
$\rho^n g_{n}({\bf r}^n)$  is proportional to
the probability density for simultaneously finding $n$ particles at
locations ${\bf r}^n\equiv {\bf r}_1,{\bf r}_2,\dots,{\bf r}_n$ within the system\cite{Ha86},
where $\rho=N/V$ is the number density. With this convention, each $g_n$ approaches
unity when all particle positions become widely separated within $V$.
Statistical homogeneity implies that $g_n$ is translationally
invariant and therefore only depends on the relative displacements
of the positions with respect to some arbitrarily chosen origin of the system, {\it i.e.},
\begin{equation}
g_n=g_n({\bf r}_{12}, {\bf r}_{13}, \ldots, {\bf r}_{1n}),
\end{equation}
where ${\bf r}_{ij}={\bf r}_j - {\bf r}_i$.

The {\it pair correlation} function $g_2({\bf r})$ is the one
of primary interest in this review. If the
system is also rotationally invariant (statistically
isotropic), then $g_2$ depends on the radial distance $r \equiv |{\bf r}|$ only, {\it i.e.},
$g_2({\bf r}) = g_2(r)$. It is important to introduce
the {\it total correlation} function $h({\bf r})\equiv g_2({\bf r})-1$, which, 
for a {\it disordered} system,  decays to zero for large $|{\bf r}|$ sufficiently rapidly.  \cite{To06b}

Such pair statistics
can be inferred from radiation scattering experiments via
the structure factor.\cite{Ha86} The {\it structure factor} $S({\bf k})$,
for an $N$-particle system  is related to the {\it collective density} variable 
\begin{equation}
{\tilde \rho}({\bf k})=\sum_{j=1}^N\exp(i {\bf k}\cdot {\bf r}_j),
\label{collect}
\end{equation}
via the expression 
\begin{equation}
S({\bf k})= \frac{|{\tilde \rho}({\bf k})|^2}{N},
\label{factor}
\end{equation}
where ${\tilde \rho}(\bf k)$ is the Fourier transform of $\rho(\bf r)$, defined
by (\ref{local}), and $i=\sqrt{-1}$. Since the structure
factor is proportional to the intensity of the scattered radiation,
it is a nonnegative quantity for all $\bf k$, {\it i.e.},
\begin{equation}
S({\bf k}) \ge 0 \qquad \mbox{for all} \quad {\bf k}.
\end{equation}
This also mathematically follows from the nonnegative form (\ref{factor}). 
In the thermodynamic limit ($N \rightarrow \infty$, $V \rightarrow \infty$
such that $\rho$ is a fixed positive constant),
the ensemble-averaged structure factor (omitting forward scattering) is defined by 
\begin{equation}
S({\bf k})=1+\rho {\tilde h}({\bf k}),
\end{equation}
where ${\tilde h}({\bf k})$ is the Fourier transform of the total correlation function
$h(\bf r)$.

The structure factor $S({\bf k})$ provides a measure of the density fluctuations
at a particular wave vector $\bf k$. To see this important property quantitatively,
consider the point pattern defined by the centers of particles in a many-body
system at number density $\rho$. Let $\sigma^2(R)$ denote the {\it number variance} of points contained within
a $d$-dimensional spherical window of radius $R$ in $\mathbb{R}^d$. It can be shown \cite{To03a}
that the number variance has the following real-space and Fourier-space representations:
\begin{eqnarray}
\sigma^2(R)&=&\rho v_1(R) \Bigg[ 1+\rho \int_{\mathbb{R}^d} h({\bf r}) \alpha(r; R) \, d{\bf r}\Bigg] \nonumber\\
&=&  \rho v_1(R)\Bigg[\frac{1}{(2\pi)^d} \int_{\mathbb{R}^d} S({\bf k})
{\tilde \alpha}({\bf k};{\bf R}) d{\bf k}\Bigg],
\label{variance} 
\end{eqnarray}
where $v_1(R)=\pi^{d/2}R^{d}/\Gamma(1+d/2)$ is the volume of the spherical
window of radius $R$, $\alpha(r;R)$ is {\it scaled intersection volume},
equal to the volume common to two  spherical windows of radius $R$ whose centers are 
separated by a distance $r$ divided by $v_1(R)$, and ${\tilde \alpha}(k;R)$ is
the corresponding Fourier transform of $\alpha(r;R)$.
The scaled intersection volume $\alpha(r;R)$ and its Fourier transform ${\tilde \alpha}(k;R)$ can be expressed explicitly
in any dimension $d$. \cite{To03a,To06b} Thus, we see that the structure factor
is directly related to the number variance at different wavelengths or, equivalently,
for different window radii. In the limit of an infinitely large window, 
the relation above yields
\begin{equation}
\lim_{R \rightarrow \infty} \frac{\sigma^2(R)}{\rho v_1(R)} = S({\bf k}={\bf 0})= 1+ \rho \int_{\mathbb{R}^d} h({\bf r}) d{\bf r}.
\label{limit}
\end{equation}
Formula (\ref{limit}) applies whether the system is in equilibrium or not. In the special
case of an equilibrium system, it is well known that infinite-wavelength density
fluctuations, as expressed by (\ref{limit}),  are proportional to the isothermal
compressibility of the system. \cite{Ha86}

For large $R$, it has been proved that $\sigma^2(R)$ cannot grow more
slowly than $\gamma R^{d-1}$ or window surface area, where $\gamma$ is a positive constant. \cite{Beck87}
We note that point processes (translationally invariant or not)
for which $\sigma^2(R)$ grows more slowly than $R^d$ ({\it i.e.}, window volume) for large $R$ are examples
of {\it hyperuniform} (or superhomogeneous) point patterns. \cite{To03a,Ga03}  Hyperuniformity
implies that the structure factor $S({\bf k})$ has the following
small ${\bf k}$ behavior:
\begin{equation}
\lim_{{\bf k}\rightarrow {\bf 0}} S({\bf k}) =0.
\label{hyper}
\end{equation}
This classification
includes all crystal structures, \cite{To03a} point patterns
associated with periodic and certain aperiodic tilings of space, \cite{Ga03,To03a,Ga04,Ga08}
one-component plasmas, \cite{Ga03,To03a} distribution of matter
 in the early Universe, \cite{Pe93,Ga01}  the ground state of superfluid helium, \cite{Fe56} 
maximally random jammed sphere packings, \cite{Do04d} and the
ground states of spin-polarized
fermions. \cite{Fe98,To08c}

\section{Inverse Methods for Crystal Ground States}
\label{method1}

We recall that a classical ground-state configuration ${\bf r}^N$ is one that minimizes
the system potential energy $\Phi_N({\bf r}^N$. Our ability to identify ground states
for a particular interaction is a highly challenging problem, \cite{Ul68,Ra91,Go05,Su05,Su06,To08a} not to mention
the even more difficult inverse problem of designing interactions to achieve
targeted ground states. Here we describe recent progress on the 
latter problem.
Because there is a vast (infinitely large) class of many-body potentials,
we begin, for simplicity,  by considering isotropic pairwise additive interactions,
{\it i.e.}, Eq. (1) reduces to the following form:
\begin{equation}
\Phi_N({\bf r}^N)  =   \sum_{i<j}^N
\varphi(r_{ij}),
\label{pair}
\end{equation}
where the {\it pair potential} $\varphi(r)\equiv \varphi_2(r)$ is a radial
function, i.e., it depends on the radial distance  $r=|{\bf r}|$. 
Although realistic interactions that operate in soft matter systems can exhibit complicated
many-particle characteristics, often a more economical description
is sought that uses at most singlet and pair effective interactions that are density dependent  to take advantage of the theoretical and computational simplifications that 
result. \cite{St02} Therefore, our starting point of pairwise additivity (in the absence of
an external field) is a practically useful approximation for
colloids and polymers, for example.

There are many open questions even for
this simple class of potentials. For instance, the limitations
of isotropic pairwise additivity for producing target structures
are not fully known and can be probed using inverse 
methods. We know that such interactions cannot
produce thermodynamically stable chiral structures with a
specified handedness; equal amounts of left-handed and
right-handed structures would result. When is anisotropy in
the potential required? An answer based on intuition from
molecular systems would fail here. For example, the diamond
crystal is thought to require directional interactions
because such structures found in Nature result from covalent bonding. In fact, 
it has recently been shown \cite{Re07a} that a diamond
structure can be created from nondirectional
interactions with strong short-range repulsions, as described in detail below.
This structure has a special status in photonics research
because a diamond crystal of dielectric spheres exhibits a
photonic band gap across the Brillouin zone. \cite{Ho90}

Two inverse optimization schemes, called the
``zero-temperature" and the ``near-melting" schemes, \cite{Re05,Re06a} 
have been devised for the
purpose of designing interactions for targeted many-particle
configurations.  Specifically, the combination
of these two optimization techniques lead to an $N$-body classical system
with particles interacting via optimized
potentials that has as its ground state ({\it i.e.}, global energy minimum
state) the corresponding 
target configuration in
a specific volume (or density) range. Unlike previous attempts
to solve this problem, this conclusion is arrived at
only after satisfying four important necessary criteria: 

\begin{enumerate}
\item lattice sums show that there is a positive pressure (or, equivalently
density) range in which
the given lattice is stable;

\item all crystal normal mode frequencies
are real at a specific density; 

\item defects (vacancies
and interstitials) are shown to cost the system energy; 

\item and  the system self-assembles in a molecular dynamics 
or (Monte Carlo) simulation
that starts above the freezing point and is slowly cooled.
\end{enumerate}

As concerns the last criterion, the system may start from an entirely random configuration
or with a layer of fixed particles to promote epitaxial
growth. Hence we make the important distinction here
between homogeneous and heterogeneous nucleation in self-assembly.
It is of course a more stringent requirement that
the desired lattice self-assemble from a random configuration (homogeneous nucleation).
Note that sufficient criteria to ensure that a ground state is exactly achieved
do not exist.

\subsection{Zero-Temperature Optimization Scheme}

In the zero-temperature optimization scheme, an optimized pair potential 
for self-assembly of a particular target
configuration at a temperature  of absolute zero is found  by
choosing a family of functions $\varphi(r;a_0,a_1,\ldots,a_n)$, parametrized
by the $a_i$'s, and then finding the values of the parameters that
lead to the most robust and defect-free self-assembly of the
target crystal for a fixed density or, preferably, a range of densities
(or, equivalently, a range of pressures).
The objective function is chosen so that  the energetic stability of a given target
crystal is maximized with respect to competitor lattices (chosen previously) subject to the condition
that the target crystal is linearly mechanically stable.
Mechanical stability is ensured for a given potential 
by establishing that that every phonon mode in the first Brillouin zone is real.
Thus, the structure is mechanically stable at zero
temperature. However, this does not preclude other structures, periodic or otherwise,
from being lower in energy than the targeted one.
Therefore, the final outcome of the zero-temperature
scheme becomes the initial potential function condition for the
near-melting optimization procedure.

\subsection{Near-Melting Optimization Scheme}

From an initial parameterized potential (final outcome of the zero-temperature
scheme),  the near-melting procedure optimizes the potential
for self-assembly at a temperature near but
below the crystal's melting point by suppressing nucleation
of the liquid phase in molecular dynamics (MD) (or Monte Carlo) simulations.
Specifically, simulations are repeatedly run at
80-95\% of the melting temperature (the temperature is chosen such
that phase-transition fluctuations do not render the calculations
inconsistent), each time calculating the Lindemann parameter, defined by
\begin{equation}
\Theta_2 = \sqrt{\frac{1}{N}\sum_i ({\bf r}_i - {\bf r}_i^{(0)})^2 -
\left( \frac{1}{N}\sum_i ({\bf r}_i-{\bf r}_i^{(0)})\right)^2},
\end{equation}
where ${\bf r}_i$ is the position of the $i^{th}$ particle (after an
appropriate amount of simulation time), ${\bf r}_i^{(0)}$ is its
initial position, and $N$ is the number of particles. The parameter $\Theta_2$ is
then taken as the objective function for a simulated annealing
calculation, and the $a_i$ are found such that
$\Theta_2$ is minimized. 

The ultimate criterion for self-assembly is
the very strong condition that the targeted ground state be observed
in a well-annealed molecular dynamics MD simulation
starting from the liquid state. The system is slowly annealed
to $T=0$ until the essentially defect-free target crystal results in reasonable
computer time. Usually only a very few defects are found
at the end of the simulation and its energy is checked to ensure
that it is higher than that of the perfect crystal.

\begin{figure}[tp]

\subfigure[]
{
 \centerline{ \includegraphics[width=8.cm,clip=]{hon_pot.eps}}

}

\subfigure[]
{
 \centerline{ \includegraphics[width=8.cm,clip=]{hon_pho.eps}}

}
\subfigure[]
{
\centerline{   \includegraphics[width=8.cm,clip=]{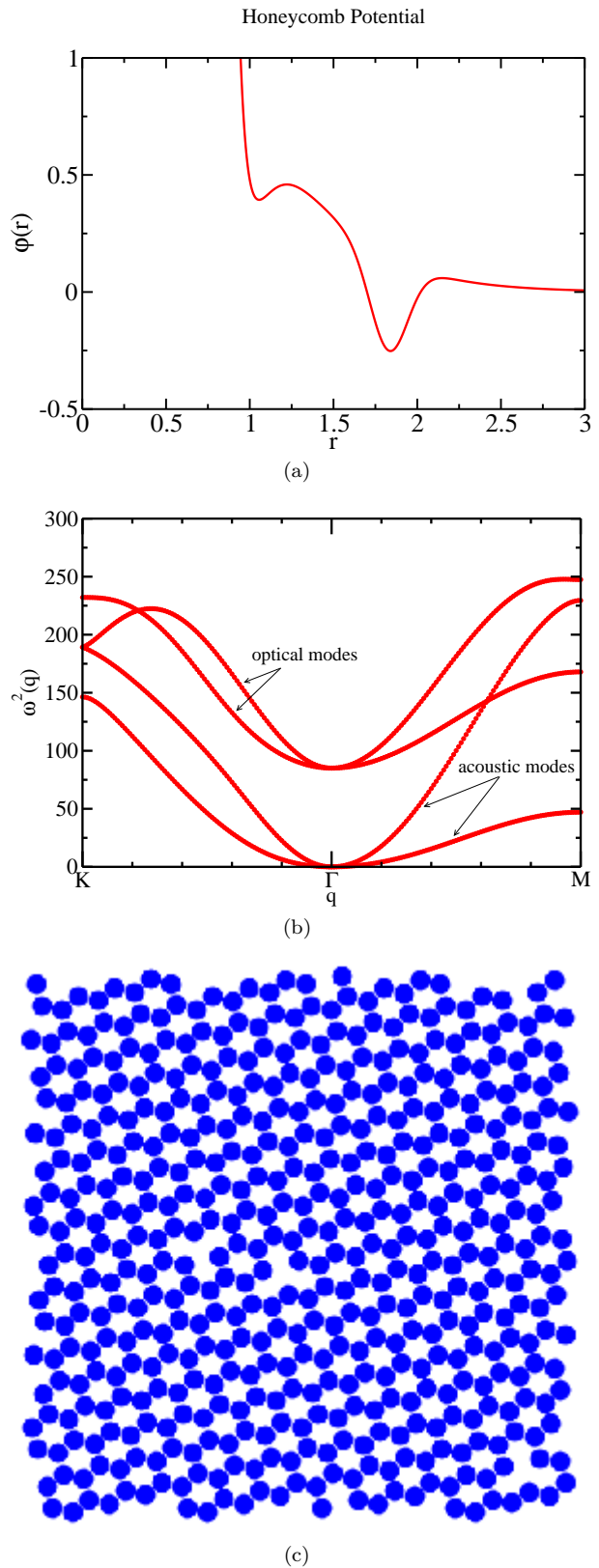}}
}
\caption{Honeycomb-crystal self-assembly as obtained
in Ref. \onlinecite{Re05}. (a) Optimized pair potential $\varphi(r)$.
Dimensionless energy and length units are defined by the axes for this potential.
(b) Phonon spectrum (frequency squared)  versus wave vector ${\bf k}$
for the optimized honeycomb-crystal potential at a specific
area equal to 1.45. The acoustic and optical branches are shown.
(c) Corresponding 500-particle annealed ground-state configuration. 
Although there are a few vacancies, these were  ``frozen
in" during annealing due to finite time of the simulation. Such vacancies
were shown to cost energy, indicating that the perfect honeycomb crystal is the true ground state.}
\label{hon_pot}
\end{figure}

\section{Optimized Isotropic Interactions for Low-Coordinated Crystal Ground States}
\label{low-coord}

Until recently, conventional wisdom presumed that low-coordinated crystal ground states require directional interactions. The aforementioned optimization schemes were tested initially
to yield optimized isotropic (nondirectional) pair potentials that spontaneously yield the 
four-coordinated square lattice and three-coordinated honeycomb lattice as ground-state structures
in two dimensions. \cite{Re05,Re06a} The latter target choice is motivated by its three-dimensional analog, the diamond lattice. Figure \ref{hon_pot} shows
the optimized honeycomb potential and corresponding
phonon spectra as well as annealed configuration
at $T=0$. It was found that as long as the salient features of the
honeycomb potential are kept (two local minima at distance
ratio $\sqrt{3}$, the first being positive and the second
negative), self-assembly is unaffected by perturbations in
the potential; {\it i.e.}, the potential is robust. This is an essential 
feature if this system is to be tested experimentally. 
We note that the functional form of the optimized ``square-lattice" potential
is simpler than that of the honeycomb crystal.

\begin{figure}[bthp]

\subfigure[]
{
 \centerline{ \includegraphics[width=7cm,clip=]{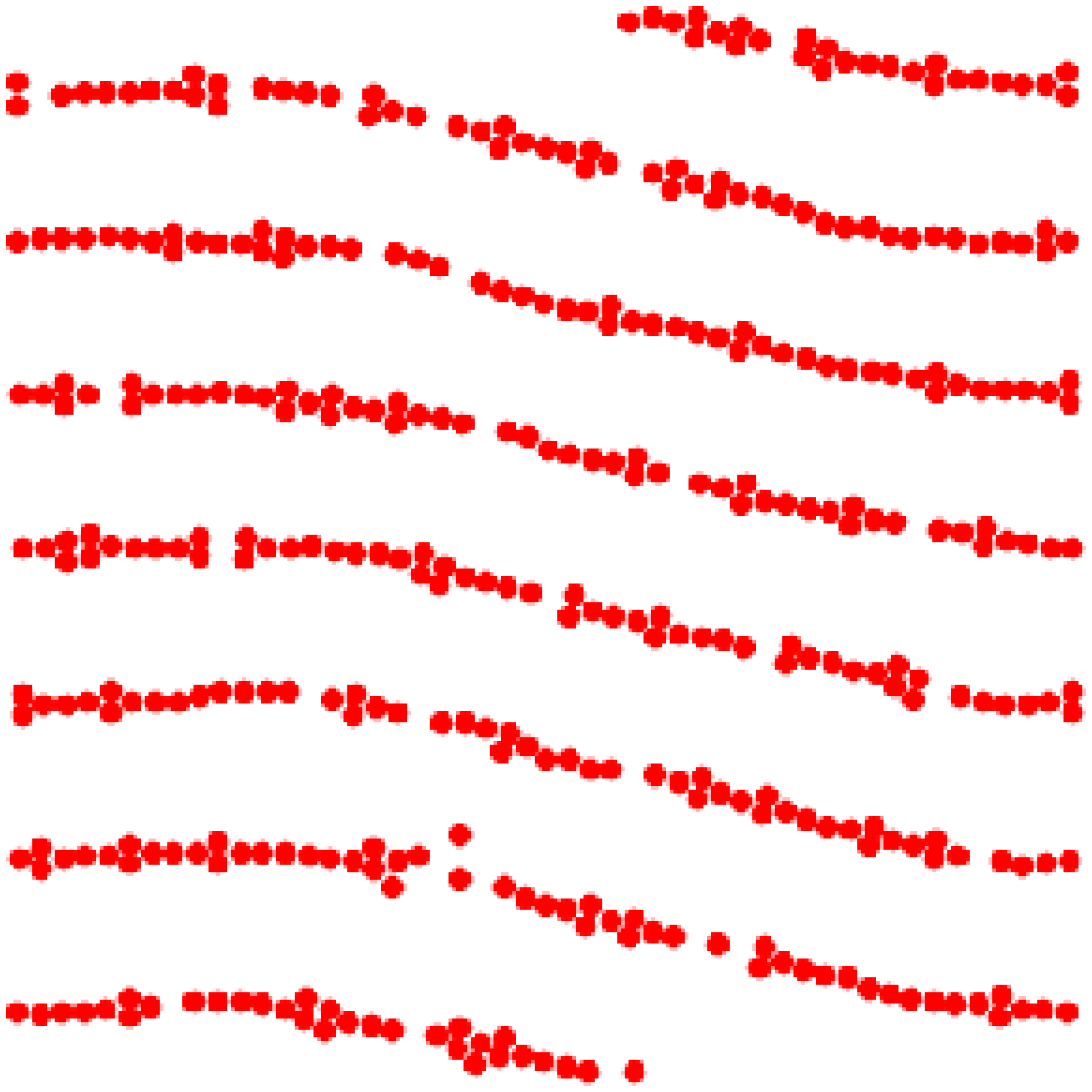}}

}
\subfigure[]
{
\centerline{   \includegraphics[width=7cm,clip=]{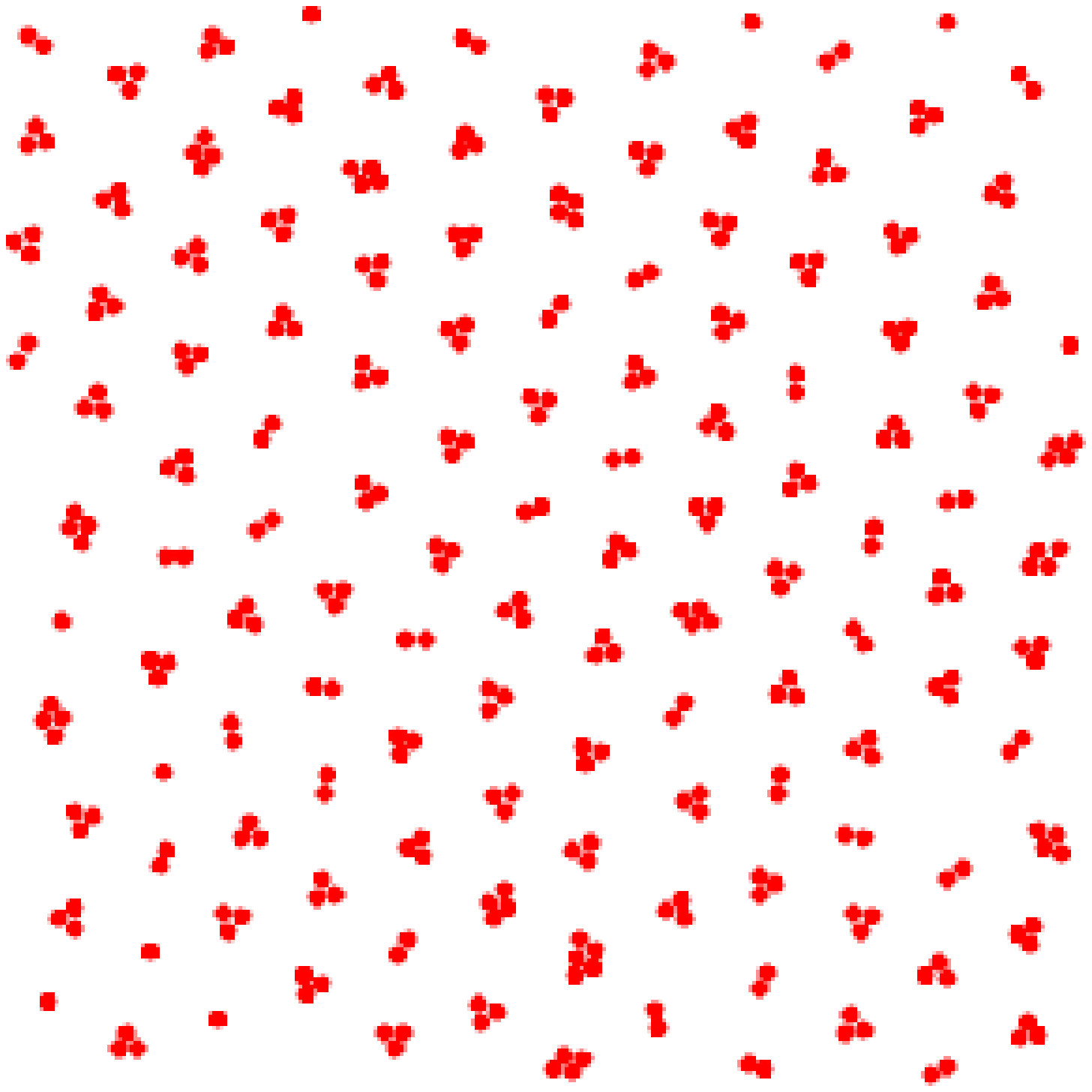}}
}
\caption{Unusual two-dimensional ground-state configurations generated from 
circularly symmetric pair potentials, as adapted from Ref.  \onlinecite{Re06a}. (a) Chain-like particle configurations. (b) Lattice of ``simplex" clusters. }
\label{clusters}
\end{figure}

In a separate work, the global phase diagram for the optimized ``honeycomb" potential 
was determined. \cite{Hy06} 
The phase diagram was obtained from Helmholtz free energies calculated using
thermodynamic integration and Monte Carlo simulations. These results showed that the honeycomb
crystal remains stable in the global phase diagram even after temperature effects are taken fully into
account. Other stable phases in the phase diagram are high- and low-density triangular phases and a fluid phase. No evidence of gas-liquid or liquid-liquid phase coexistence was found.

In order to test the limitations of circularly symmetric potentials in two dimensions,
pair potentials were devised that yielded structurally
anisotropic chain-like arrays as well as lattices of compact clusters, \cite{Re06a}
as shown in Fig. \ref{clusters}.
These structures are reminiscent of ``colloidal wires"  and ``colloidal clusters"
found experimentally by the authors of Refs.  \onlinecite{Stell04} and \onlinecite{Mano03}, respectively.
Interestingly, we see that structural anisotropy (colloidal wires) can counterintuitively
be achieved with isotropic interactions with  the so-called ``five-finger" potential. 
\cite{Re06a} This potential cannot be built in the lab with current technology, 
but it shows that isotropic potentials have perhaps
more flexibility than one would immediately think. It is
also very possible that a much simpler isotropic potential could allow for a
similar structure to assemble.

These two-dimensional results were extended to the self-assembly of low-coordinated
three-dimensional crystals with isotropic pair interactions, including 
the determination of  an optimized pair potential whose classical ground state is the
simple cubic lattice and which is functionally simple enough to synthesize
in the laboratory. \cite{Re06b} The same investigation
reported optimized isotropic potentials 
that yield the body-centered-cubic
and simple hexagonal lattices (planes of triangular lattices stacked
directly on top of one another), which provide other examples of non-close-packed structures that can be
assembled using only isotropic pair interactions.

\begin{figure}

\subfigure[View 1]
{
 \centerline{ \includegraphics[width=6cm]{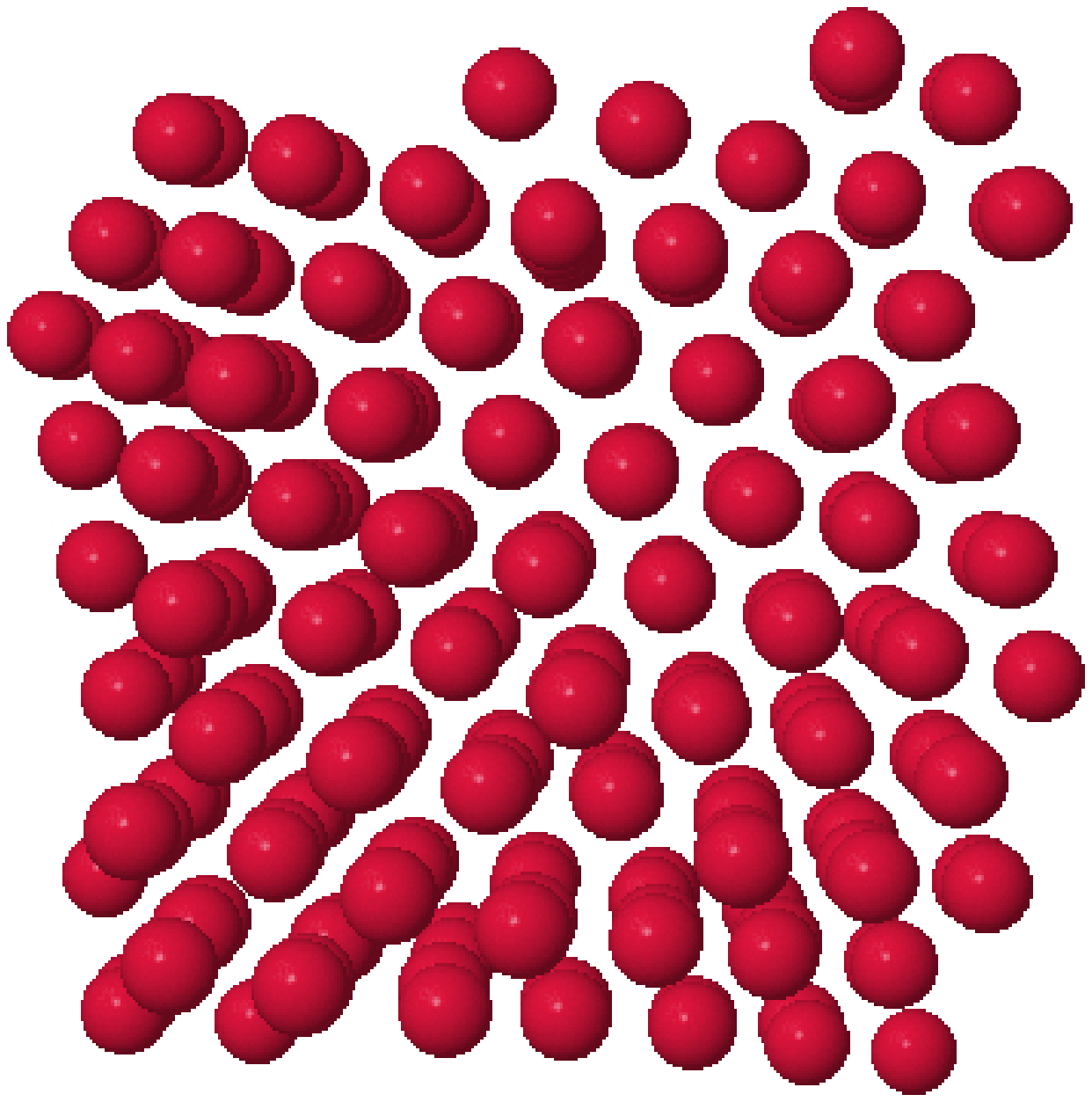}}

}
\subfigure[View 2]
{
\centerline{   \includegraphics[width=6cm]{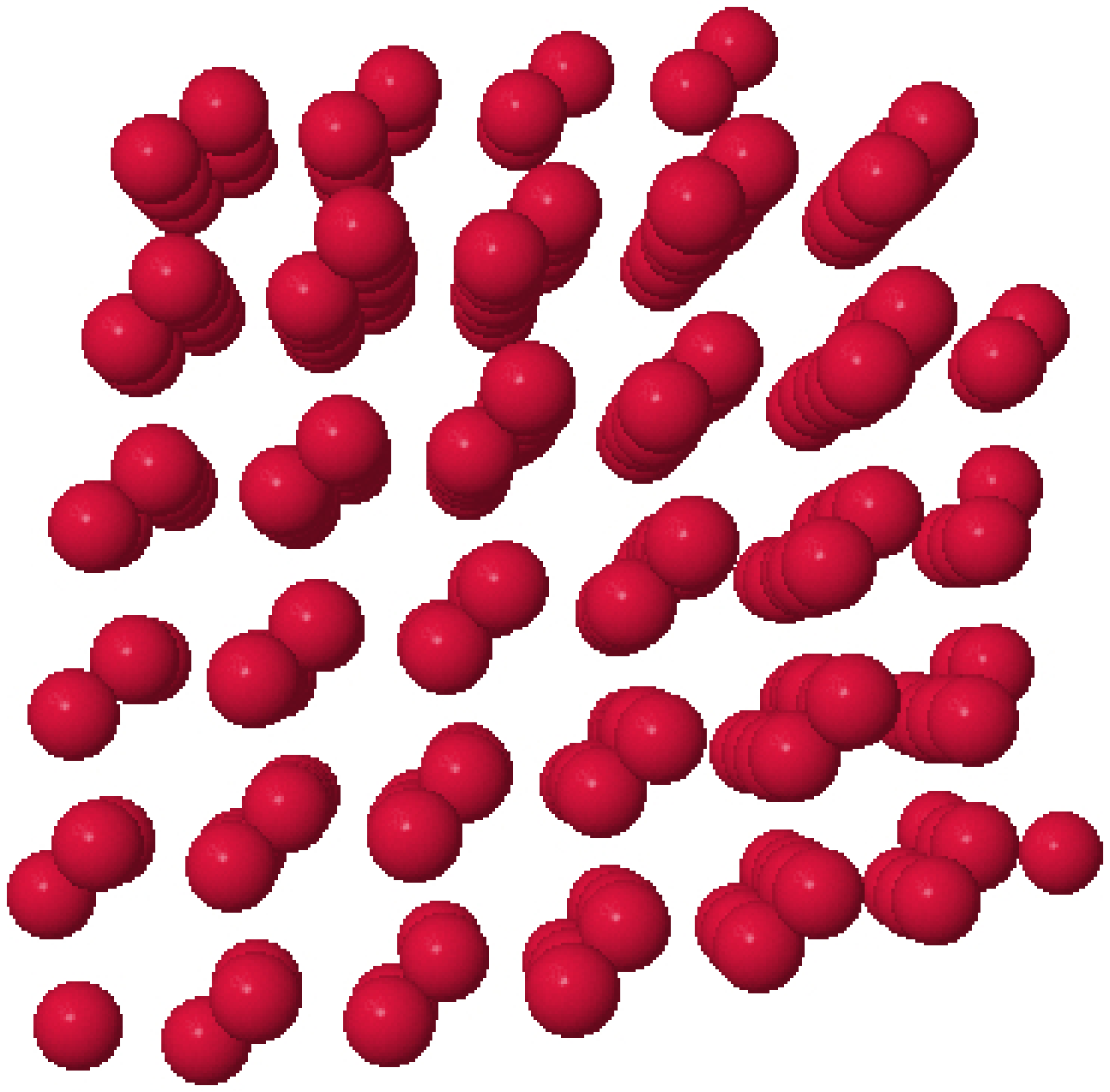}}
}
\subfigure[View 3]
{
 \centerline{  \includegraphics[width=6cm]{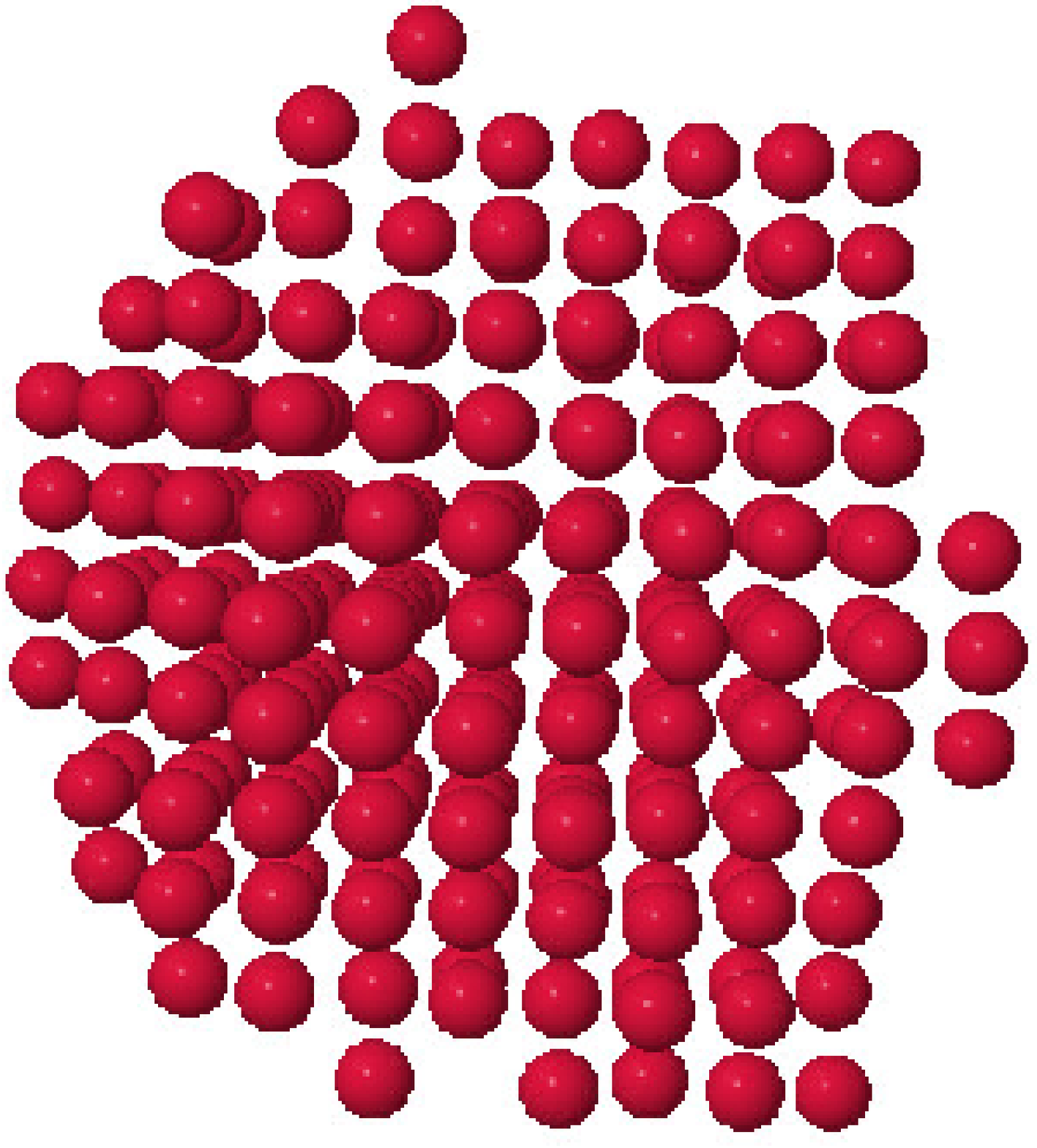}}
}

\caption{Results of an MD simulation for   216 particles interacting via
  the optimized isotropic ``diamond" potential showing  self-assembly into a perfect diamond
  configuration, as  in Ref. \onlinecite{Re07a}. One configuration is 
shown from three different  viewpoints, which  clearly show that the
  result is the diamond crystal.}
\label{diamond}
\end{figure}

Optimized isotropic pair-interaction
potentials with strongly repulsive cores have been obtained that cause the 
tetrahedrally coordinated diamond and wurtzite lattices
to stabilize, as evidenced by lattice sums, phonon spectra, positive-energy defects, 
and self-assembly in classical molecular dynamics simulations. \cite{Re07a}
Figure \ref{diamond} depicts one self-assembled diamond-crystal configuration shown from three different
  viewpoints.
Finding such a potential via inverse
methods is a highly nontrivial problem, since the
diamond crystal is extremely close in structure to the tetrahedrally-coordinated {\it wurtzite} crystal in particular.    Given
the functional form of the potential, the pressure (or volume) was
tuned very precisely to find a small stability range for the diamond structure,
and under such conditions, simulations readily demonstrated its
self-assembly.  
These results challenge conventional thinking that such open lattices can
only be created via directional covalent interactions observed in nature
and adds to our fundamental understanding of the nature of the solid state.

Note that it has been shown that an
isotropic pair potential that models star polymer systems 
has a region of phase stability that favors the diamond crystal. \cite{Wa99} However, this
potential, in contrast to the one reported in Ref. \onlinecite{Re07a}, possesses a soft core,
which would be difficult to synthetically produce using colloids.

\section{Inverse Optimization Methods for Disordered Ground States}
\label{method2}

Collective density variables $\rho({\bf k})$ [cf. (\ref{collect})] have proved to be
useful tools in the study of static and dynamic
phenomena occurring in many-body systems. \cite{Pe58,Fa91} 
More recently, the collective-coordinate approach  has been used to generate crystalline as
well as noncrystalline classical ground states for bounded or ``soft"
interactions using numerical optimization techniques in two and three dimensions. 
\cite{Uc04b,Uc06a,Ba08}
Soft interactions possess great importance in soft-matter systems,
such as colloids, microemulsions,  and polymers. \cite{Ru89,Wa99,La00,Ml06,Gl07}

We begin by briefly reviewing the basic description
of the collective-coordinate approach for $N$ interacting particles
in a $d$-dimensional cubic box of side length $L$ and volume $\Omega$ under periodic boundary conditions.
The corresponding infinite set of wave vectors is given by
\begin{equation}
   {\bf k} = (\frac{2\pi n_{1}}{L}, \frac{2\pi n_{2}}{L},\ldots,\frac{2\pi n_{d}}{L})\;,
\label{Equation2}
\end{equation}
where the $n_{i}$ ($i=1,2,\ldots,d$)  are positive or negative integers, or zero.
The structure factor, defined by (\ref{factor}), can be rewritten as follows:
\begin{displaymath}
   S({\bf k}) = \frac{|\rho({\bf k})|^2}{N}=1 + \frac{2}{N} C({\bf k}),     
\end{displaymath}
where the real collective density variable $C({\bf k})$ is subject to the following constraints:
\begin{eqnarray}
    C({\bf 0}) &=& \frac{1}{2} N (N-1)  \\
    C({\bf k}) &=& C(-{\bf k})  \\
    -\frac{1}{2} N \leq C({\bf k}) &\leq& \frac{1}{2} N (N-1) \hspace{10pt} ({\bf k} \neq 0).
\label{Equation5}
\end{eqnarray}
The lower bound on $C({\bf k})$ arises because the structure factor $S({\bf k})$
must be nonnegative.

Let us assume that the total potential energy $\Phi_N$ is pairwise additive and therefore
given by (\ref{pair}). Suppose furthermore that the pair potential 
$\varphi(r)$ has a Fourier transform ${\tilde \varphi}({\bf k})$:
\begin{eqnarray}
    {\tilde \varphi}({\bf k}) &=& \int_{\Omega} \varphi({\bf r}) \exp(i{\bf k} \cdot {\bf r}) dr, \\
    \varphi({\bf r}) &=& \Omega^{-1} \sum_{{\bf k}} {\tilde \varphi}({\bf k}) \exp(-i{\bf k}
         \cdot {\bf r}),
\label{Equation8}
\end{eqnarray}
where in the last expression the summation covers the entire set of ${\bf k}$'s. Then it is
straightforward to show that the total potential energy for the $N$-particle system can be exactly
expressed in the following manner in terms of the real collective density variables
\begin{equation}
   \Phi_N({\bf r}^N) = \Omega^{-1} \sum_{{\bf k}} {\tilde \varphi}({\bf k}) C({\bf k}).
\label{Equation9}
\end{equation}
Consider pair interactions whose
transform ${\tilde \varphi}({\bf k})$ is nonnegative
radial function $f(k) \ge 0$ ($k=|{\bf k}|$) with compact support, {\it i.e.},
\begin{equation}
{\tilde \varphi}(k)=  f(k)\Theta(K-k),
\label{f}
\end{equation}
where
\begin{equation}
\Theta(x) =\Bigg\{{0, \quad x<0,\atop{1, \quad x \ge 0,}}
\label{heaviside}
\end{equation}
is the Heaviside step function. We see that if the $C({\bf k})$ is driven to
its minimum value $-N/2$ for $|{\bf k}| < K$,  then that configuration must
be a classical ground state of the system, the absolute
minimum of $\Phi$. Thus, density fluctuations for those
${\bf k}$'s such that $|{\bf k}| < K$ are completely suppressed,
{\it i.e.}, the structure factor $S({\bf k})=0$ for $|{\bf k}| < K$.
Note that for the form (\ref{f}), the corresponding
real-space pair potential $\varphi(r)$ will be an oscillating
potential. However, there are choices for $f(k)$ one can make, especially
for purposes of experimental realizability, that can appreciably
dampen the amplitudes and range of the real-space interactions.

Although the number of collective variables is infinite, the $N$-particle system possesses only $dN$
configurational degrees of freedom, where $d$ is the Euclidean space dimension.  Consequently, 
it is unreasonable to suppose (barring special circumstances) that generally all $C({\bf k})$'s could be
independently controlled.  However,  it is possible, as illustrated below, to specify 
simultaneously a number of the collective variables equal to a significant
fraction of $dN$. We denote by $\chi$  the ratio of the constrained degrees of freedom to
the total number of degrees of freedom. As $\chi$ increases to cover
larger and larger numbers of the wave vectors, and consequently
having an impact on a larger and larger fraction of the total
degrees of freedom, the result for the classical ground state is
far from obvious. It is clear that if $\chi$ is a fraction of order
unity, the ground state is periodic, which has been established. \cite{Fa91,Uc04b,Su05,Uc06a,Ba08}

However, the more interesting cases involve disordered ground states
({\it i.e.}, configurations that possess no long-range order), which
arise for a range of $\chi \in [0,\chi_{max}]$, provided
that $\chi_{max}$ is sufficiently small. \cite{Uc04b,Uc06a,Ba08}
Our primary interest here are in the disordered, degenerate ground states
that can be produced by the collective-density approach.
Such systems have the remarkable property of 
being able to self-assemble into one of the numerous degenerate
disordered configurations when slowly cooled to absolute zero.

For any given choice of $N$ and $K$, the 
numerical procedure utilizes a random number generator to create
an initial configuration of the particles inside the 
hypercubic box. This starting point typically produces a large
positive value of the system potential energy $\Phi_N$. The next
step involves use of an optimization procedure,
such as the conjugate gradient method or a more sophisticated
technique, \cite{Uc06a} to
seek a particle configuration that yields the absolute minimum
value of $\Phi_N$.

This numerical optimization technique has been employed  to generate 
two-dimensional classical ground-state particle configurations
with the simple transform choice $f(k)=1$ in (\ref{f}), {\it i.e.}, the pure
unit step function, which is zero for $k>K$. \cite{Uc04b}  The resulting investigation
distinguished three structural regimes as the number of constrained
wave vectors is increased ({\it i.e.}, as $\chi$ is increased) - disordered, wavy crystalline, and
crystalline regimes.

The aforementioned collective-coordinate procedure has been generalized to those
cases in which $C({\bf k})$ is constrained to be some
target value $C_{0}({\bf k}) \ge 0$ for ${\bf k} \in {\bf Q}$, 
where $\bf Q$ represents the finite set of ${\bf k}$'s for
which a number of the  collective density variables can simultaneously be specified.
Of course, each $C_{0}({\bf k})$ must lie in the range specified by 
inequalities of (\ref{Equation5}). Then consider the following non-negative objective function: 
\begin{equation}
   \Phi_N({\bf r}^N) = \sum_{{\bf k} \in \textbf{Q}} 
   {\tilde \varphi}({\bf k}) [C({\bf k}) - C_{0}({\bf k})]^{2}.   
\label{Equation7}
\end{equation}    
If $\Phi_N$ is interpreted as a potential energy of interaction for the $N$ point particles, then 
it can be shown that it represents intrinsic two-body, three-body and four-body
interaction potentials operating in the system.  If
classical ground-state configurations for the $N$ particles subject to that potential exist for which $\Phi_N = 0$,
then those configurations necessarily attain the desired target values of the collective variables.

This generalization of the collective coordinate approach
was applied in three dimensions. \cite{Uc06b} In particular, multi-particle 
configurations were generated for which $S({\bf k}) \propto
|{\bf k}|^{\alpha}$, $|{\bf k}| \leq K$, and $\alpha =$ 1, 2, 4, 6, 8, and 10.  
The case $\alpha = 1$ is relevant for the Harrison-Zeldovich model of 
primordial density fluctuations of the early Universe, \cite{Pe93,Ga01} 
superfluid helium, \cite{Fe56} maximally random jammed 
sphere packings, \cite{Do05d} and spin-polarized
fermions.\cite{Fe98,To08c}  This analysis also provides specific examples of interaction potentials whose
classical ground states for finite-sized
systems are configurationally degenerate  and disordered.

Employing this collective-coordinate numerical optimization procedure, ground-state configurations of interacting particle systems in the first three space dimensions have been constructed  so that the scattering of radiation exactly matches a prescribed pattern for a set of wave vectors. \cite{Ba08} It is demonstrated that the
constructed ground states are, counterintuitively, disordered ({\it i.e.}, possess no long-range order) in the {\it infinite-volume} limit. Three classes of configurations with unique radiation scattering characteristics
were studied: (i)``stealth'' materials, which are transparent to incident radiation at certain wavelengths; (ii)``super-ideal'' gases, which scatter radiation identically to that of an ensemble of ideal gas configurations for a selected set of wave vectors; and (iii)``equi-luminous'' materials, which scatter radiation equally intensely for a selected set of wave vectors.  

Although stealth materials and super-ideal gases are subsets of equi-luminous
materials, we use this term to refer to materials that scatter radiation more
intensely relative to an ideal gas. These materials that scatter radiation much more intensely
than an ideal gas for a set of wave vectors have enhanced density fluctuations and show local
clustering similar to polymers and aggregating colloids. \cite{Scha89,Brink90} 
With the collective-coordinate inverse procedure, the degree of clustering can be
imposed by tuning the scattering characteristics for certain wavelengths.

\begin{figure}[tp]

\subfigure[]
{
 \centerline{ \includegraphics[width=8.65cm,clip=]{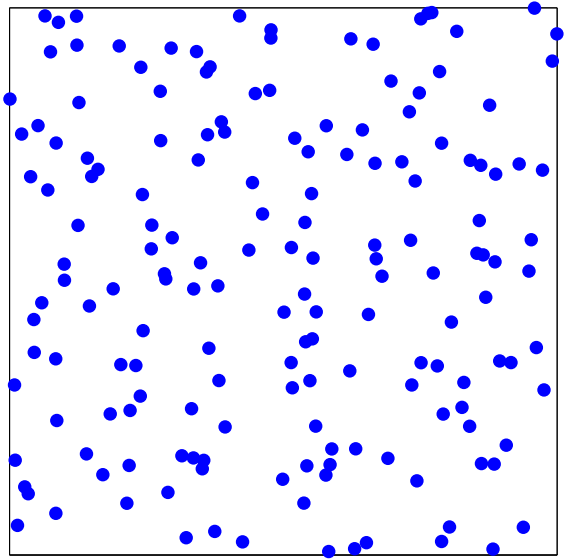}}

}
\subfigure[]
{
\centerline{   \includegraphics[width=8cm,clip=]{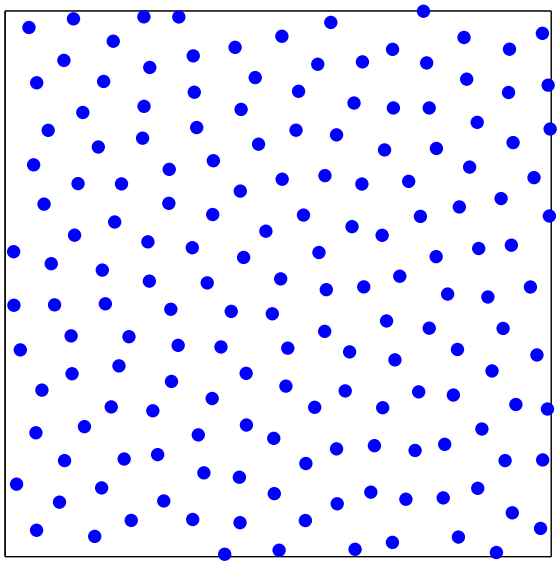}}
}
\caption{Stealth particle patterns of 168 particles in two
dimensions, as adapted from Ref. \onlinecite{Ba08}: (a) $\chi$ = 0.04167, (b) $\chi$ = 0.20238.  
Both systems are disordered but at higher $\chi$,
particles tend to repel one another to a greater degree. The potential energy
was minimized to within 10$^{-17}$ of its global minimum.}
\label{stealth}
\end{figure}

For purposes of illustration, disordered ``stealth" configurations are depicted
in two dimensions in Figure \ref{stealth} for 168 particles 
for two selected values of $\chi$. At the lowest $\chi$ considered, the
configuration is seen not  to have strong spatial correlations. At highest
$\chi$ value reported, the particles develop an exclusion shell about their centers
but the system still does not possess any long-range order. A system size study
was carried out that revealed no long-range order
when extrapolated to the infinite-volume limit.

\section{Duality Relations for Classical Ground States}
\label{Duality}

The determination of the classical ground states  of
interacting many-particle systems (global minimum energy configurations) is an outstanding problem in condensed-matter
physics and materials science. \cite{Ul68,Ra91}
While classical ground states are readily produced
by slowly freezing liquids in experiments and computer simulations, our theoretical
understanding of them is far from complete.

Much of the progress to rigorously identify ground states for given interactions has been 
done for lattice models, primarily in one dimension. \cite{Ra91} The solutions
in $d$-dimensional Euclidean space $\mathbb{R}^d$ for $d \ge 2$
are considerably more challenging.  For example, the ground state(s) for the
well-known Lennard-Jones potential in $\mathbb{R}^2$ or $\mathbb{R}^3$ are
not known rigorously (although many computer
simulations support the conclusion that 
the hexagonal close-packed crystal is its ground state). 

Soft (bounded) interactions are easier to treat
theoretically as evidenced by recent progress in our understanding
of ground states of this class of potentials in $\mathbb{R}^2$ and $\mathbb{R}^3$.
\cite{Uc04b,Su05,Uc06a,Su06,Ba08}
Moreover, as we noted earlier, such interactions possess great importance in 
a variety of soft-matter systems. \cite{Ru89,Wa99,La00,Ml06,Gl07}

Nonetheless, new theoretical tools are required to make further progress.
Duality relations that link the energy of configurations associated with a
class of soft pair potentials to the corresponding energy of the dual 
(Fourier-transformed) potential have recently been derived. \cite{To08a}
These duality relations enable one to use information 
about ground states of short-ranged potentials
to draw new conclusions about the nature
of the ground states of long-ranged potentials and vice versa.
Among other results, they also have led to the identification of
unusual one-dimensional systems with ground-state ``phase transitions" 
and can be employed to make computational searches
for ground states more efficient.

Before discussing these duality relations, which take the form
of two theorems, we introduce some notation.
Let $U({\bf r}^N)$ be  twice the total potential energy per particle
in an $N$-particle system with pairwise interactions, {\it i.e.},
\begin{equation}
U({\bf r}^N)=\frac{1}{N}\sum_{i, j} \varphi(r_{ij}),
\end{equation} 
where $\varphi(r)$ is a radial pair potential function and $r_{ij}=|{\bf r}_j-{\bf r}_i|$.
A {\it classical ground-state} configuration  is one that minimizes $U({\bf r}^N)$.
We consider those stable radial pair potentials $\varphi(r)$ that are bounded and
absolutely integrable and call such functions admissible.
Thus, the corresponding Fourier transform ${\tilde \varphi}(k)$ in $d$ dimensions
at wave number $k$ exists. 
We recall that in a Bravais lattice $\Lambda$, the space $\mathbb{R}^d$ can be geometrically divided into identical regions called {\it fundamental cells}, each of which contains 
one particle center.\cite{footnote} We denote the reciprocal Bravais lattice by $\tilde \Lambda$.
If the Bravais lattice $\Lambda$ has density $\rho$, then its reciprocal
lattice $\tilde \Lambda$ has density ${\tilde \rho}=\rho^{-1}(2\pi)^{-d}$.
\bigskip

\noindent{\bf Theorem 1.}
{\sl If an admissible pair potential $\varphi(r)$ has a Bravais lattice $\Lambda$
ground-state structure at number density $\rho$, then we have the following duality relation
for twice the minimized energy per particle $U_{min}$:
\begin{equation}
\varphi(r=0)+ {\sum_{{\bf r} \in \Lambda}}^{\prime} \varphi(r) = \rho {\tilde \varphi}(k=0)+ 
\rho {\sum_{{\bf k} \in {\tilde \Lambda}}}^{\prime} {\tilde \varphi}(k),
\label{duality}
\end{equation}
where the prime on the sum denotes the zero vector should be omitted, ${\tilde \Lambda}$ denotes the reciprocal Bravais lattice,
and ${\tilde \varphi}(k)$ is the dual pair potential, which automatically satisfies the
stability condition, and therefore is admissible.
Moreover, twice the minimized energy per particle ${\tilde U}_{min}$
for any ground-state structure of the dual potential ${\tilde \varphi}(k)$, is
bounded from above by the corresponding real-space {\it minimized} quantity $U_{min}$
or, equivalently, the right side of (\ref{duality}), {\it i.e.},
\begin{equation}
{\tilde U}_{min} \le U_{min}=\rho {\tilde \varphi}(k=0)+ \rho  {\sum_{{\bf k} \in {\tilde \Lambda}}}^{\prime} {\tilde \varphi}(k).
\label{bound}
\end{equation}
Whenever the reciprocal lattice  ${\tilde \Lambda}$ at {\it reciprocal lattice density}
${\tilde \rho}=\rho^{-1}(2\pi)^{-d}$ is a ground state of ${\tilde \varphi}(k)$,
the inequality in  (\ref{bound}) becomes an equality.
On the other hand, if an admissible dual potential ${\tilde \varphi}(k)$ has
a Bravais lattice ${\tilde \Lambda}$ at number density ${\tilde \rho}$,
then
\begin{equation}
U_{min} \le {\tilde U}_{min}={\tilde \rho} \varphi(r=0)+ {\tilde \rho}  
{\sum_{{\bf r} \in {\Lambda}}}^{\prime} \varphi(r),
\label{bound2}
\end{equation}
where equality is achieved when the real-space ground state is the lattice $\Lambda$
reciprocal to ${\tilde \Lambda}$.}
\smallskip

Whenever equality in relation (\ref{bound}) is achieved,
then a ground state structure of the dual potential ${\tilde \varphi}(k=r)$
evaluated at the real-space variable $r$ is the Bravais lattice ${\tilde \Lambda}$
at density ${\tilde \rho}=\rho^{-1}(2\pi)^{-d}$. 
Theorem 1 leads to another theorem (both of which
are proved in Ref. \onlinecite{To08a}) concerning phase coexistence.
\bigskip

\noindent{\bf Theorem 2.}
{\sl Suppose that for admissible potentials there exists a range of densities over which
the ground states are side by side coexistence
of two distinct crystal structures whose parentage are two
different Bravais lattices, then the strict inequalities in
(\ref{bound}) and (\ref{bound2}) apply at any density
in this density-coexistence interval.} 
\smallskip

Note that the ground states referred to in Theorem 2 are not only non-Bravais lattices, they are not even periodic. The ground  states are side-by-side coexistence of two crystal domains
whose shapes and relative orientations are complicated functions of $\rho$.

On account of the ``uncertainty principle" for Fourier pairs,
a non-localized (long-ranged) potential $\varphi(r)$ has a corresponding
localized (compact) dual potential ${\tilde \varphi}(k)$.
Similarly, a localized (compact) potential $\varphi(r)$ has a corresponding
non-localized (long-ranged) dual potential ${\tilde \varphi}(k)$.
This property of Fourier pairs and the duality relations of Theorem 1
enable one to use information about ground states of short-ranged potentials
to draw new conclusions about the nature
of the ground states of long-ranged potentials and vice versa.
In particular, three different classes of admissible
potential functions have been considered: (1){\it compactly}
supported functions (such as the ones employed in the collective-coordinate
approach discussed in Section \ref{method2}); (2) {\it nonnegative}
functions; and (3) {\it completely monotonic} functions.

For purposes of illustration, we discuss here in some detail,
the application of Theorem 1 to the class potential
functions that have been used in the collective-coordinate
approach  reviewed in Section \ref{method2}.
Recently, the ground states
corresponding to a certain class of oscillating real-space potentials $\varphi(r)$ as defined
by the family of Fourier transforms  with compact support
such that ${\tilde \varphi}(k)$ is positive for $0 \le k < K$
and zero otherwise have been studied. \cite{Uc04b,Su05} Clearly, ${\tilde \varphi}(k)$
is an admissible pair potential.  In Ref. \onlinecite{Su05},
it was shown that in three dimensions the corresponding  real-space potential $\varphi(r)$,
which oscillates about zero,  has the body-centered
cubic (bcc) lattice as its unique ground state at the real-space density
$\rho=1/(8\sqrt{2}\pi^3)$ (where we have taken $K=1$). Moreover, it was
demonstrated \cite{Su05} that for densities greater than $1/(8\sqrt{2}\pi^3)$,
the ground states are degenerate such that the face-centered cubic (fcc),
simple hexagonal (sh), and simple cubic (sc) lattices are ground states
at and above the respective densities $1/(6\sqrt{3}\pi^3)$, $\sqrt{3}/(16\sqrt{2}\pi^3)$, and
$1/(8\sqrt{2}\pi^3)$.

Because all of the aforementioned ground states are Bravais
lattices, the duality relation (\ref{duality}) can be applied
to infer the ground states of real-space potentials
with compact support. Specifically, application of the duality theorem in $\mathbb{R}^3$
and the results of Ref. \onlinecite{Su05} enables us to conclude that for the real-space potential
$\varphi(r)$ that is positive for $0 \le r < D$ and zero otherwise,
the fcc lattice (dual of the bcc lattice) is the unique ground state
at the density $\sqrt{2}$ and the ground states are degenerate such that the bcc, sh
and sc lattices are ground states at and below the respective densities $(3\sqrt{3})/4$, $2/\sqrt{3}$, and
$1$ (taking $D=1$). Specific examples
of such real-space potentials, for which the ground
states are not rigorously known, include the ``square-mound" potential 
[$\varphi(r)=\epsilon >0$ for $0 \le r <1$ and zero otherwise] and
the ``overlap" potential $\alpha(r;D/2)$, \cite{To03a} equal to the intersection
volume of two $d$-dimensional spheres of diameter $D$ whose
centers are separated by a distance $r$ divided by the
volume of a sphere (discussed in Section \ref{background}), and thus has support in the interval $[0,D)$.   Moreover, any structure, periodic or not,
in which the nearest-neighbor distance is greater than unity is
a ground state.

\begin{figure}[bthp]
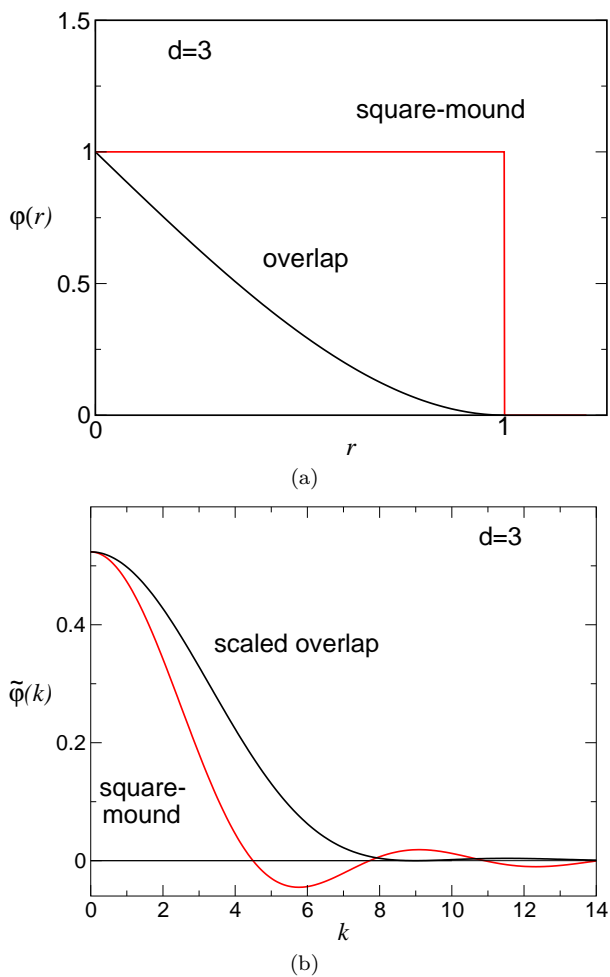


\subfigure[]
{
 \centerline{ \includegraphics[width=8cm,clip=]{compact-support.eps}}

}
\subfigure[]
{
\centerline{   \includegraphics[width=8cm,clip=]{delocalized.eps}}
}
\caption{(a) The localized square-mound potential
[$\varphi(r)=\epsilon=1$ for $0 \le r <1$ and zero otherwise] and overlap potential [$\varphi(r)=1-3r/2+r^3/2$ for $0 \le r <1$ and zero otherwise]
in $\mathbb{R}^3$. (b) The  delocalized dual square-mound potential ${\tilde \varphi}(k)=\pi^{3/2}J_{3/2}(k)/(2k)^{3/2}$
multiplied by $\pi^3/6$ and dual overlap potential ${\tilde \varphi}(k)=6\pi^2 J^2_{3/2}(k/2)/k^3$.}
\label{compact}
\end{figure}

Importantly, at densities corresponding to nearest-neighbor distances
that are less than unity,  the possible ground-state structures is considerably
more difficult to ascertain. For example, it has been argued in Ref. \onlinecite{Ml06} (with good reason) that real-space
potentials whose Fourier transforms oscillate about zero will
exhibit polymorphic crystal phases in which the particles
that comprise a cluster sit on top of each other. The square-mound potential
is a special case of this class of potentials and the fact
that it is a simple piecewise constant function
allows for a rigorous analysis of the clustered ground states
for densities in which the nearest-neighbor distances
are less than the distance at which the discontinuity
in $\varphi(r)$ occurs.

The duality relations have also led to the identification of a one-dimensional system that exhibits
an infinite number of ``phase transitions" at $T=0$  from
Bravais to non-Bravais lattices over the entire
density range as well as  a conjecture regarding the ground states
of purely repulsive monotonic potentials. \cite{To08a}
Moreover, inequalities (\ref{bound}) and (\ref{bound2}) provide a computational tool
to estimate ground-state energies or eliminate
candidate ground-state structures as obtained from annealing simulations.

The Gaussian potential is a special case of a purely repulsive monotonic potential,
and is a useful interaction to model polymer systems. \cite{La00,Ml06}
The phase diagram of such systems in various spatial dimensions
has recently been investigated \cite{Za08} in order to understand the 
effect of dimensionality, apply the aforementioned duality relations,
and to test a conjecture of Ref. \onlinecite{To08a} concerning
completely monotonic potentials. The Gaussian potential is an example
of the class of potentials in which both the real-space and dual
potentials are nonnegative functions. The authors of Ref. \cite{Ml06}
have argued that such systems display re-entrant
melting with an upper freezing temperature.

Elsewhere, corresponding duality relations for potential
functions that also include three-body and higher-order
interactions will be derived \cite{To08b}.

\section{Construction of Configurations with Target Pair Correlations}
\label{realize}

The subject of atomic and molecular distribution functions
has enjoyed a long and rich history. However, not surprisingly for a scientific area so characterized
by intrinsic complexity, some deep problems of incomplete
understanding still persist.

One such open question concerns “realizability” of a given candidate pair correlation function
$g_2({\bf r})$, namely, whether it actually
represents the pair correlation 
of some many-particle configuration at number density
$\rho >0$. This is called the {\it realizability} problem.\cite{To02c,To03a} Several necessary
conditions that must be satisfied by the candidate are known, including nonnegativity of 
$g_2({\bf r})$ and its associated structure factor $S({\bf k})$, as well as constraints on implied local density fluctuations. \cite{Ya61}
It has recently come to light that a positive $g_2$ at a positive $\rho$ must
satisfy an uncountable number of necessary and sufficient
conditions for it to correspond to a  realizable  point process. \cite{Cos04,Ku07}
However, these conditions are very difficult (or, more likely, impossible)
to check for arbitrary dimension. In other words, given $\rho$
and $g_2$, it is difficult to ascertain if there
are some higher-order functions $g_3, g_4, \ldots$ for which these one- and two-particle
correlation functions hold.

To shed light on the realizability problem, 
a simple one-dimensional lattice model, with single-site occupancy, and 
nearest-neighbor exclusion has been investigated. \cite{St04}
The following results were obtained: 
(a) pair correlation realizability over a nonzero density range, (b) violation of
the Kirkwood superposition approximation for $g_3$, and (c) inappropriateness of the so-called
``reverse Monte Carlo" method that uses a candidate pair correlation function as a means to suggest typical many-body configurations. Note that Chayes and Chayes \cite{Cha84}
proved that for any pair correlation function (meeting mild conditions) that is
derivable from an $N$-body Hamiltonian, there always exists a unique
``effective" two-body potential  that produces the same pair correlation function
(but generally not the higher-body correlation function $g_3, g_4, g_5$, etc.).
This theorem has been successfully applied to polymer solutions to obtain
effective pair interactions from $g_2$ \cite{Bo01a,Bo01b}.

Elsewhere, so-called iso-$g_2$ processes were studied in the equilibrium regime.
These consist of a sequence of equilibrium
many-body systems that have different number densities  but share, at a given temperature, the same ``target" pair correlation function. In other words, in these processes,
density-dependent interactions identically cancel the usual density variation of 
many-body pair correlation functions. \cite{St01a,Sa02,St05}
Target pair correlation functions studied include the unit step function
as well as the zero-density limit of the square-well potential
(for which $g_2(r)=\exp[-\beta \varphi(r)]$).
Formal density expansions for effective pair potentials 
were derived with this iso-$g_2$ property, showing how successive terms in that expansion can be determined iteratively. Explicit
results through second density order have been obtained for two types of ``target" pair
correlation functions, and the conditions under which
realizability can be attained were explored. \cite{St05}

In order to explore and gain insight into the basic statistical 
geometric features of random sphere packings, the notion of a 
$g_2$-invariant process was introduced. \cite{To02c}
A {\it $g_2$-invariant process} is one in which a given nonnegative pair correlation
$g_2({\bf r})$ function remains invariant as density varies for all ${\bf r}$ over the range of
densities
\begin{equation}
0 \le \rho \le \rho_*.
\end{equation}
The terminal density $\rho_*$ is the maximum achievable density
for the $g_2$-invariant process subject to satisfaction of
the known necessary conditions on the pair correlation function.
The determination of the terminal density for various
forms of $g_2$ that putatively correspond to a sphere packing
has been solved using numerical and analytical optimization
techniques. \cite{To06b,To02c,To06d,Sc08}

To test whether such $g_2$'s at terminal density
$\rho_*$ are indeed realizable by sphere packings, 
stochastic optimization techniques originally, developed
to construct material microstructures with 
targeted lower-order correlation functions, \cite{Ri97a,Ye98a,Cu99}
were employed. \cite{Uc06a,Cr03}
In a construction algorithm, an initial configuration
of particles evolves such that the final configuration possesses a set
of targeted correlation functions up to some ``cut-off" distances. This is done by choosing
the objective function to be a ``squared error"
involving the set of targeted correlation functions.
The evolving configurations are induced by minimizing this objective 
function via a stochastic optimization procedure.

\begin{figure}[bthp]

\subfigure[]
{
 \centerline{ \includegraphics[width=8cm,clip=]{Figure2a.eps}}

}
\subfigure[]
{
\centerline{   \includegraphics[width=8cm,clip=]{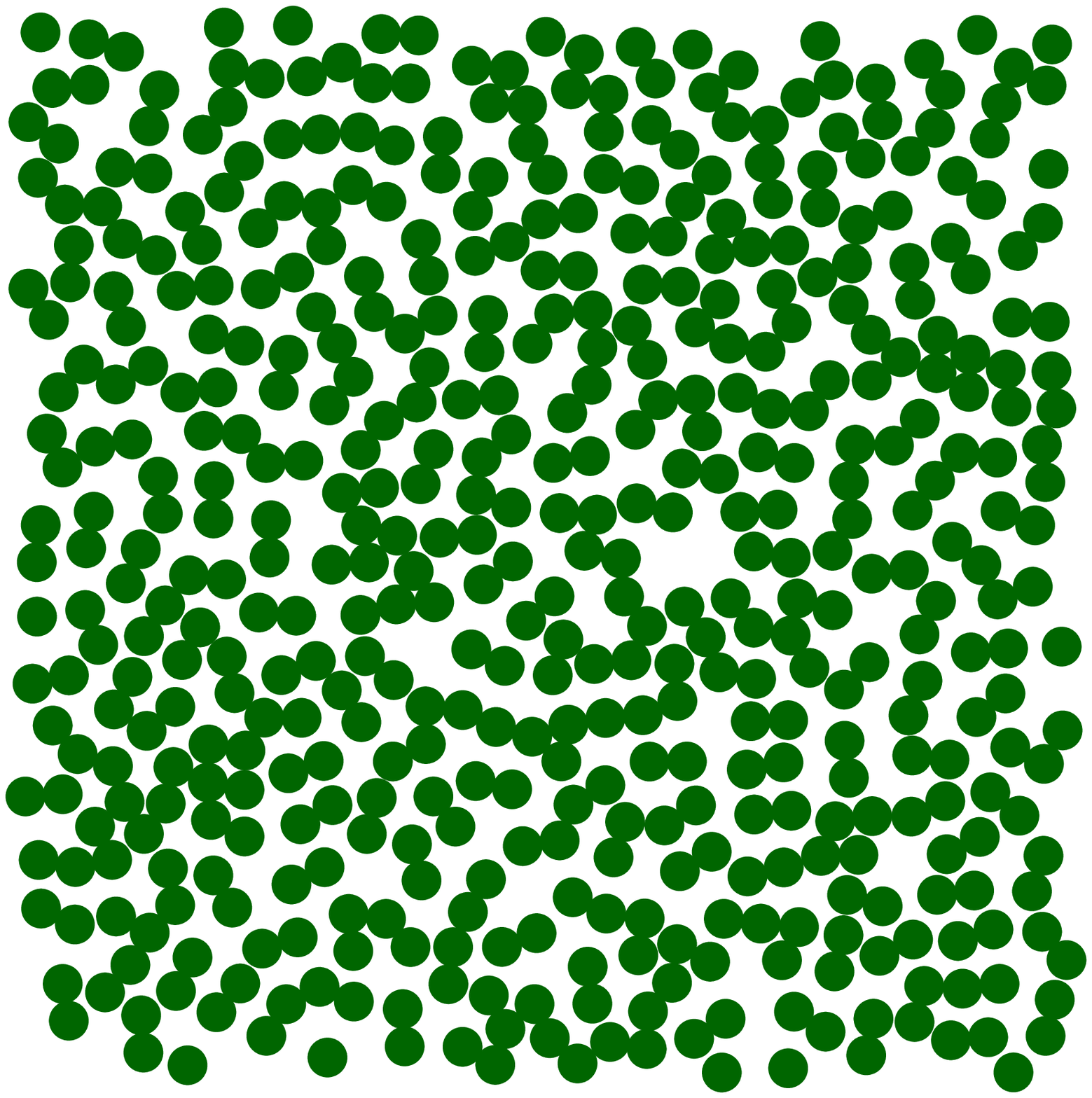}}
}
\caption{(a) Graph of the target pair correlation function
$g_2(r)$: Dirac $\delta$ function plus a step function. 
(b) A two-dimensional configuration of 500 particles that realizes
this targeted form for $g_2(r)$ up to a dimensionless distance
of $r/D=2.5$, as adapted from Ref. \onlinecite{Uc06a}. The configuration consists of only dimers at the terminal
packing fraction $\phi_{*}$ = $0.5$ with an average contact value $Z$ = 1.0.}
\label{fig:IConstruction}
\end{figure}

For example, for the case of $d$-dimensional packing of congruent spheres
of diameter $D$ in which the pair correlation function
is taken to be
\begin{equation}
g_2(r)=\frac{Z}{\rho s_1(r)}\delta(r-D) + \Theta(r-D),
\end{equation}
where $Z$ represents the average contact value per sphere 
and $s_1(r)=d\pi^{d/2}r^{d-1}/\Gamma(1+d/2)$ is the surface area of a $d$-dimensional sphere,\cite{To02c} it was found
that the terminal packing fraction $\phi_*$ (fraction of space
covered by the spheres) and the associated average contact number $Z_*$
are given by
\begin{equation}
\phi_*= \frac{d+2}{2^{d+1}}, \qquad Z_*= \frac{d}{2}.
\end{equation}
Numerical evidence suggests that such a pair correlation 
is achieved by a single sphere packing configuration for any $d \ge 2$. \cite{Uc06a}
 Such a pair correlation function 
Figure \ref{fig:IConstruction} shows a realization
of such a packing in two dimensions.
Of course, in any simulation, pair distances must binned
and sampled up to some cut-off distance. Note that for a sufficiently
large system, the targeted correlation for a single configuration
approaches that of one obtained from an ensemble of configurations
by ergodicity.

Because the realizability problem is far from being solved,
it remains an active area of research. For example, 
it has been conjectured that any radial, nonnegative pair correlation
function characterized by a hard-core, which decays sufficiently rapidly
to unity,  is realizable by a translationally invariant disordered sphere
packing in $d$-dimensional Euclidean space for asymptotically large $d$ 
if and only if $S(k) \ge 0$.\cite{To06a} Although there is mounting
evidence to support this conjecture,\cite{To06a,To06b,Sc08} a proof of it is
a great challenge.

\section{Designing Isotropic Pair Potentials for Targeted Bulk Properties}
\label{bulk}

Inverse methods have been recently devised to optimize interactions of 
many-particle systems to
achieve targeted novel bulk properties. To illustrate the interesting possibilities,
we discuss three specific target examples in some detail: the thermal expansion
coefficients and Poisson's ratio.

\subsection{Thermal Expansion Coefficients}

Control of thermal expansion properties of materials is of
technological importance due the need for structures to withstand
ambient temperature variations.  In the technological realm, materials with zero thermal expansion
(those that do not expand or contract upon heating) can aid in the
longevity of space structures, bridges and piping
systems. \cite{Si97,Gi97c} Materials
with very large thermal expansion coefficients could function as
actuators, and those with negative thermal expansion  coefficients may
be of use as thermal fasteners.\cite{Si96}

Negative thermal expansion (NTE) behavior, a well-known but unusual phenomenon in many-particle
systems, has been observed only in multi-component materials with open
unit cell structures in which the bonding of component particles is
highly directional.  Perhaps the most common example of a solid exhibiting NTE is that of
ice, which contracts upon melting into liquid water. \cite{Fl70} Another
example of a material that undergoes NTE is zirconium tungstate, $Zr
W_2 O_8$, which exhibits this behavior for an extremely large
temperature range, namely $0.3K$ through $1050K$. \cite{Mar96}

\begin{figure}[bthp]
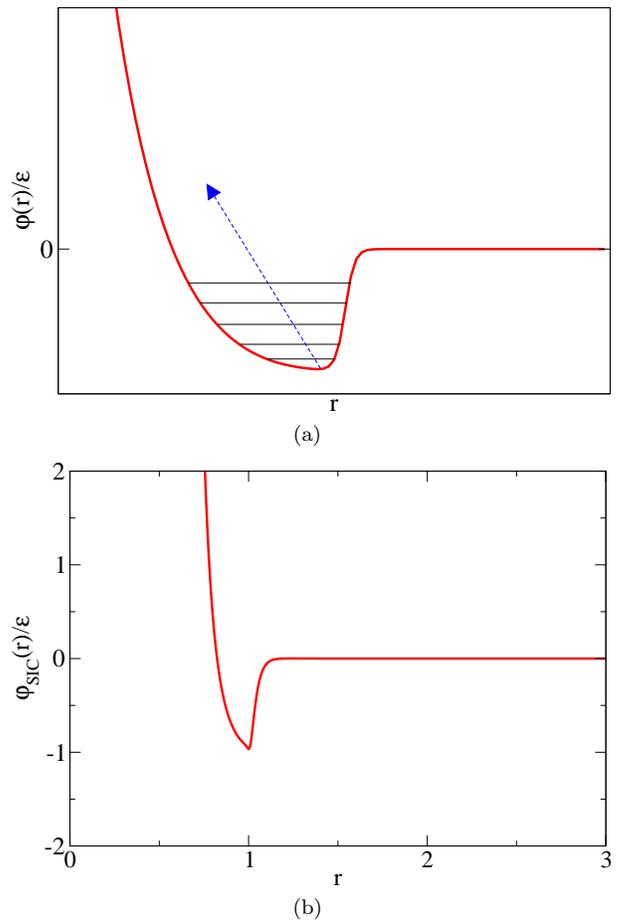


\subfigure[]
{
 \centerline{ \includegraphics[width=8cm,clip=]{schematic.eps}}

}
\subfigure[]
{
\centerline{   \includegraphics[width=8cm,clip=]{potential-concocted.eps}}
}
\caption{(a) Schematic depiction of an isotropic pair potential 
(scaled by the well depth $\epsilon$) with a
softened interior in its basin of attraction following
Ref. \onlinecite{Re07b}. Thermal fluctuations cause
the average nearest-neighbor distance to decrease, resulting in an overall
contraction of the system upon heating. (b)  The
optimized ``softened interior core" (SIC)  potential,
as adapted from Ref. \onlinecite{Re07b}, has NTE behavior over
a wide range of temperatures.}
\label{expansion}
\end{figure}

An isotropic interaction potential has been optimized that gives rise to negative thermal expansion (NTE) behavior
   in equilibrium many-particle systems in the solid state in both two and three dimensions over a wide temperature and pressure
   range (including zero pressure). \cite{Re07b}  Although such anomalous behavior is well-known in materials with directional
   interactions ({\it e.g.}, zirconium tungstate), this is the first time that NTE behavior has been
   established to occur in the solid state of single-component many-particle systems for isotropic interactions. (Note that NTE has been shown to occur in a two-dimensional {\it fluid}
with isotropic interactions. \cite{Ja99,Ca03}) 
Moreover,  it was established that a sufficient condition for a potential to
give rise to a system with NTE behavior is that it exhibits a
softened interior core within a basin of attraction (as depicted
schematically in part (a) of Fig. \ref{expansion}).
   Using an optimization procedure to find a potential that yields a
   strong NTE effect and constant-pressure Monte Carlo simulations, 
it was shown that as the temperature was increased, 
the ``softened interior core" potential
[part (b) of Fig. \ref{expansion}], the system exhibited negative,
   zero, and then positive thermal expansion before melting (in both two and three dimensions).

\subsection{Poisson's Ratio}

Another interesting target bulk property is the Poisson's
ratio $\nu$. In particular, it is desired to optimize interactions
to achieve negative Poisson's ratio (NPR), the so-called ``auxetic" materials \cite{Ba93,Ba03}.
When such materials are stretched in a particular direction, they expand in an orthogonal direction.  Auxetic
behavior is a counterintuitive material property that has been observed only in a handful
of elastically isotropic materials that often have intricate structures and characteristic lengths much larger
than an atomic bond length, such as foams \cite{La87} and other cellular 
materials. \cite{Mi92,Si96-1,Gi97,Xu99} Auxetic materials have a great deal of technological potential;
for example, they can be used as strain amplifiers. \cite{Ba03} If 
auxetic materials are used as a matrix in the manufacture of miniature sensors based on
piezoceramic composites, the range of operating frequencies of a piezoelectric
transducer is widened and the sensitivity of the device is increased.
\cite{Gi97b}  They can also be used as mechanical
components of microelectromechanical systems, and as transducing structures, shock
absorbers and fasteners. \cite{Xu99}

\begin{figure}
\centerline{\includegraphics[width=4in]{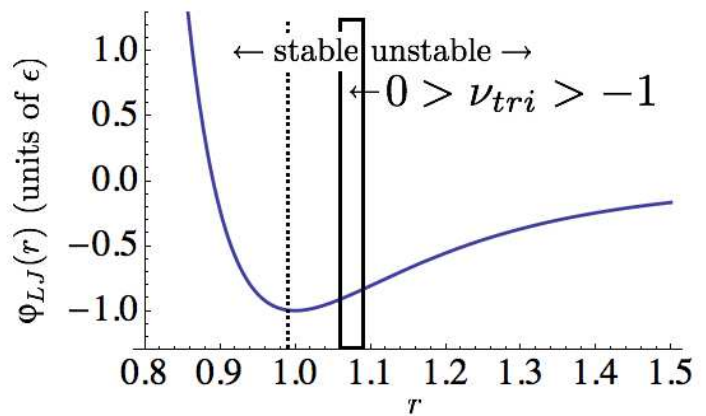}}
\caption{Region of lattice constants (indicated by the rectangular box) for
which the Poisson's ratio is negative in a triangular lattice, using the Lennard-Jones
interaction potential, $\varphi_{LJ}$, as adapted from Ref. \onlinecite{Re08c}  Pressure is positive to the left of the dotted line and
negative to the right; thus, auxetic behavior only occurs at {\it negative pressure}.  To the right of
the rectangular box, the lattice becomes unstable. }
\label{lennard-jones-plot}
\end{figure}

It has been recently found that under {\it tension} ({\it i.e.}, negative pressure), many-body two- and three-dimensional systems with
isotropic two-body interaction potentials can have a negative Poisson's ratio in the crystal
phase as long as certain linear equalities and inequalities involving the interaction potential
$\varphi(r)$ are satisfied. \cite{Re08c}  This is an unexpected result, since it describes an inherently anisotropic
behavior that arises from isotropic interactions; indeed, most previously discovered auxetic
materials exhibit complex, carefully designed anisotropic interactions.  
This can be shown  to be
the case at zero temperature for the elastically isotropic triangular lattice in two dimensions,
and for the fcc lattice in three dimensions, which, surprisingly,
can also be made to be elastically isotropic.  One can show that in the former case, the
simple Lennard-Jones potential can give rise to auxetic behavior (see Fig. \ref{lennard-jones-plot}).
In the three-dimensional case, auxetic behavior is exhibited even
when the elastic constants are constrained such that the material is elastically isotropic. 
Finding auxetic behavior over a wide range in temperature and pressure is a challenging
\emph{optimization} problem that has yet to be addressed.

This analysis suggests that auxetic behavior only occurs 
in crystals under the nonequilibrium condition of  negative pressures 
when the system contains only pair interactions and is elastically isotropic.
Such auxetic materials may potentially be experimentally produced using 
synthetic techniques that rely
on kinetic effects; examples include tempered glass, \cite{TemperedGlass} and even
colloidal crystals. \cite{ColloidalCrystalsUnderTension}  
However,  a three-body potential has been
devised that yields NPR behavior in close-packed two- and three- dimensional lattices by construction at zero temperature and
positive pressure. \cite{Re08c}  In order to produce this behavior, the potential has a built-in energy
cost associated with deforming the equilateral triangles in the two-dimensional triangular
lattice and the three-dimensional close-packed lattices. The interested
reader is referred to Ref. \onlinecite{Re08b} for the explicit
form of this three-body potential.

\section{Future Work and Conclusions}
\label{future}

In this section, we discuss future directions and close with
concluding remarks.

\subsection{Interaction Potentials for Targeted Configurations
at Positive Temperature}

Most of the inverse techniques reviewed here were
directed toward obtaining ground state ($T=0$) structures.
However, the same methods can be extended to treat
many-particle configurations at positive temperature.
For example,  an ability to control the formation of point, line, and planar defects
of crystals under various growth conditions at positive temperature is highly desirable.
The required interactions to achieve representative
amorphous target structures, including equilibrium liquids
at positive temperature and low-temperature glasses,
is another interesting application.

\subsection{Interaction Potentials for Targeted Multicomponent Systems}

\begin{figure}[bthp]
 \centerline{ \includegraphics[width=8cm,clip=]{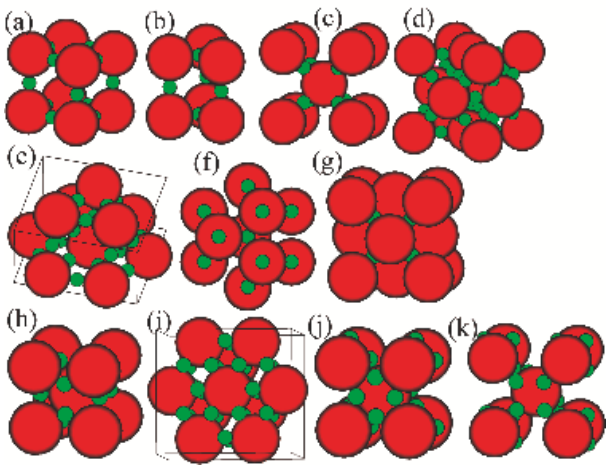}}
\caption{ Theoretically predicted stable binary crystals of oppositely charged colloids with
different stoichiometries, as obtained from Ref. \onlinecite{Hyn06}
(with the permission of the authors).}
\label{antti}
\end{figure}

It is  straightforward  to extend the zero-temperature and near-melting optimization schemes \cite{Re05,Re06a} to multicomponent systems. The parameter space,
which now includes species composition and effective particle
size ratios, becomes much larger than the single-component instance,
and therefore one must be careful in selecting the family
of potential functions that must be optimized as well
as the target structures. In order to make the search manageable, one could
limit the choice of potential functions to those  that are consistent with
interparticle interactions found in colloidal systems.
In addition to hard-sphere-like interactions, these
include  long-range repulsive, short-range attractive and averaged dipolar
interactions. It has recently been 
shown  that the electrostatic interaction between oppositely charged particles,
which are long-range attractive interactions, can result in a rich class of stable ionic colloidal crystals, \cite{Hyn06} as illustrated Figure \ref{antti}.
Motivated by this remarkable investigation,
one can imagine optimizing a family of potential functions based on 
such interactions that target an even broader class
of crystal structures.

\subsection{Anisotropic Interactions}

We have seen that  there exists 
nontrivial families of radial pair potentials for which 
interesting targeted structures are the stable low-temperature forms.  Consequently, it was not 
necessary in principle to call upon angle-dependent or non-additive 
interactions to form such nonconventional  lattices.
However, anisotropic pair interactions offer greater flexibility
to achieve targeted structures and therefore provides a new direction
to apply our inverse methods.

 Recently, a new generation of colloidal particles
with chemically or physically patterned surfaces has been designed and synthesized in the
attempt to manipulate the valency of the colloidal particles. \cite{Mano03,Ch05,van06} This synthesis effort aims to
generate ``superatoms" ({\it i.e.}, atoms at the nano and microscopic length
scales) in order to reproduce and extend traditional collective molecular 
behavior to larger length scales; thus,
opening the new ﬁeld of ``supra-particle" colloidal physics. 

One simple way to model such interactions is via ``patchy" particles, {\it i.e.}, particles with discrete, attractive interaction sites at prescribed locations on the particle surface.
Molecular simulations have been carried
out to investigate the self-assembly of patchy
particles. \cite{Zh04,Wi07,Bia08} Chains, sheets, rings, icosahedra, square pyramids, tetrahedra, and twisted and staircase structures have been obtained through suitable design of the surface pattern of patches.  Patchy particles represent a new class of building blocks for the fabrication of colloids 
with unique structural characteristics. 

Thus, it would be highly desirable to
optimize patchy particles interactions to achieve low-coordinated
crystal structures, amorphous structures, and quasicrystals.
Again, this can be accomplished by appropriate simple extensions
of the zero-temperature and near-melting optimization schemes
that were originally implemented for isotropic interactions.

\subsection{Inverse Optimization Methods for Novel Targeted Bulk Properties}

A full-blown and general optimization scheme that can be used
to find optimized interactions over a large family of potential
functions for a given
set of bulk properties over a wide range of conditions has yet to be devised.
For example, optimizing interactions in a many-particle system
so that it exhibits auxetic   behavior over a wide range in temperature and pressure is a challenging problem.
One path toward the general goal is to formulate a methodology that incorporates
a set of bulk properties in the objective function in the same
spirit as has been done for topology optimization
of composite materials, \cite{Si96,Si97,Hy01,To02d}
but in a molecular dynamics simulation. 
Specifically, the objective function can either be the
bulk property itself (which is extremized) or
a squared ``error" function involving a targeted bulk
property (which is minimized) during a molecular dynamics
simulation. The simulation would start from some 
initial configuration and randomly distributed
velocities for a initial parameterized potential. At fixed time intervals, the objective
function would be computed and then  the parameters of the
potential  updated according to some optimization
routine ({\it e.g.}, simulated annealing). 
This procedure would then be iterated until the objective function is extremized.

An intriguing set of target materials are those that exhibit
``inverse melting." \cite{Gre00} 
Inverse melting is a first-order phase transition involving the crystal and liquid, 
but with a reversal from conventional melting in that addition of heat to the liquid, at constant pressure, causes that liquid to freeze into a crystalline solid.  As a result of this reversal, the crystal has higher entropy than the isotropic liquid with which it coexists.  This is a rare phenomenon, but real-world examples exist. For example, the transition itself forms the basis of the 
zone-refining method for purification. \cite{Pf66} 
Inverse melting has been studied as a ``forward" problem using
the Gaussian core potential model. \cite{Fe03} However, devising optimized
interactions to make such unusual macroscopic behavior as robust
as possible over a wide range of conditions has heretofore not been considered.

\subsection{Toward Experimentally Realistic Interactions}

An important component of future research should be
the development of robust potentials (even if not optimal)
for targeted structures and bulk properties that 
can be synthesized experimentally with colloids or
other soft-matter systems.
It is clear that there is wide class of target structures and bulk properties  
that can be achieved with pairwise additive potentials,
both isotropic and anisotropic. In future research, it will be highly
desirable to determine, when possible, 
the pair interactions that can be either be synthesized experimentally with colloids
using current technology ({\it e.g.}, depletion, screening length,
dipolar interactions, etc.) or can be done so in the near future.
The latter could serve as a challenge to experimentalists.

Real interactions in many-particle materials at nondilute concentrations are necessarily nonadditive, {\it i.e.}, intrinsic three-body and higher-order interactions beyond pair interactions [explicitly given in Eq. (1)] are inevitable.\cite{Bo01b,St02} Thus, it is 
crucial to determine how the effective pair potentials that result from the inverse
approach correspond to the many-body interactions that arise in actual colloidal systems. This important  problem has received little attention in the literature. It has been shown that effective pair interactions that 
approximate nonadditive potentials are in fact density dependent and hence one must be
careful in carrying out the resulting statistical mechanics. \cite{St02} Guided by experiments,
one can determine 
the real two-body and three-body interactions that together mimic the effective pair potential required to 
achieve the targeted many-particle configurations using both theoretical
techniques and molecular dynamics simulations. This will require continual feedback between theory and experiment.

\subsection{Incorporating Dynamics}

The dominant theme of this review article concerned the determination of potentials
that spontaneously create target structures under equilibrium
or near-equilibrium circumstances. A conjugate
kinetic problem also exists, in which selection among alternative
irreversible scenarios (involving distinct dynamical evolutions)
itself becomes a tool for selecting among alternative structural
outcomes.  The full potential of self-assembly to control and manipulate the 
structure of materials at the microscopic and nanoscopic level
cannot be realized without a deeper understanding of
nonequilibrium processes at those length scales.
For example, a recently developed model demonstrates 
this point by showing how the irreversible collisions 
in particle suspensions that generally produce diffusive chaotic dynamics can also cause 
the system to self-organize to avoid future collisions. \cite{Chaik08}
This can lead to a  self-organized non-fluctuating quiescent state, with a dynamical phase transition separating it from fluctuating diffusing states. 
This investigation and many other nonequilibrium studies, too numerous to list here, 
provide exciting glimpses into the future of self-assembly.
Inverse optimization techniques that exploit the dynamics of many-particle systems 
to achieve self-assembly has yet to be developed and should offer greater flexibility for novel material design.

\subsection{Conclusions}

Although in their infancy, the inverse approaches reviewed here have already 
shown a capability for controlling self-assembly to an 
exquisite degree. Indeed, future applications could  revolutionize the manner 
in which materials are designed and fabricated, especially
if there is continual feedback between theory and experiment.
There are recent examples in which output from material optimization
studies have been combined with experiments to produce
novel materials or material components. \cite{Do04b,Ma05,To05,Fa07}
These inverse methods have led to a deeper fundamental understanding of the mathematical relationship between the collective structural behavior of many-body systems and their interactions. For example, we have seen that low-coordinated crystal structures,
chain-like arrays, and layered structures  do not require directional 
interactions for self-assembly. \cite{Re05,Re06a,Re06b,Re07a} Although
soft matter  with some of the interactions reviewed in this article
cannot be synthesized with current technology, other optimized interactions 
described here that yield either novel structures or bulk properties 
are rather standard or could easily made in the laboratory \cite{Re06b,Re07b,Re08c}.
For practical purposes, it will be important that future research be directed
toward producing optimized interactions with the constraint
that they are experimentally achievable. 
We envision being able to  ``tailor'' potentials that result in novel
materials with varying degrees of
disorder, thus extending the traditional idea of self-assembly to
incorporate not only crystals but  amorphous and  quasicrystal structures.
\bigskip

\noindent{\bf  Acknowledgments}
\smallskip

This review article would not have been possible without my numerous
collaborators.  In particular, I am very grateful
to Mikael Rechtsman, Obioma Uche and Robert Batten,
who were instrumental in the interaction optimization
work. I express special thanks to Frank Stillinger, who made
major contributions to this research and was my co-author
on a majority of the papers highlighted in this article.
I am grateful to Antti-Pekka Hynninen for supplying the image
for Fig. \ref{antti}. I thank the Institute for Advanced Study for its hospitality during his stay
there. This work was supported by the Office of Basic Energy Sciences,
U.S. Department of Energy, under Grant DE-FG02-04-ER46108.


\end{document}